\documentclass{vldb}

\usepackage[hidelinks]{hyperref}
\usepackage{subcaption}
\usepackage{graphicx}
\usepackage{xcolor}
\usepackage{xspace}
\usepackage{array}
\usepackage[noend]{algpseudocode}
\usepackage{dblfloatfix}
\usepackage{balance}
\usepackage{algorithm}

\newcommand{\reffigure}[1]{Figure~\ref{#1}}
\newcommand{\refsection}[1]{Section~\ref{#1}}

\newcommand{\reftable}[1]{Table~\ref{#1}}

\newcommand{\refappendix}[1]{Appendix}

\newcommand{\zap}[1]{}
\newcommand{\btree}{B$^+$-tree\xspace}
\newcommand{\btrees}{B$^+$-trees\xspace}

\vldbTitle{Efficient Data Ingestion and Query Processing for LSM-Based Storage Systems (Extended Version)}
\vldbAuthors{Chen Luo, Michael J. Carey}
\vldbDOI{10.14778/3303753.3303759}
\vldbVolume{12}
\vldbNumber{5}
\vldbYear{2019}

\begin{document}
	
\title{Efficient Data Ingestion and Query Processing for LSM-Based Storage Systems (Extended Version)}

\numberofauthors{2}

\author{
	\alignauthor
	Chen Luo\\
	\affaddr{University of California, Irvine}\\
	\email{cluo8@uci.edu}
	\alignauthor
	Michael J. Carey\\
	\affaddr{University of California, Irvine}\\
	\email{mjcarey@ics.uci.edu}\\
}
\maketitle

\begin{abstract}
In recent years, the Log Structured Merge (LSM) tree
has been widely adopted by NoSQL and NewSQL systems
for its superior write performance.
Despite its popularity, however, most existing work has focused on LSM-based key-value stores with only a single LSM-tree;
auxiliary structures, which are critical for supporting ad-hoc queries, have received much less attention.
In this paper, we focus on efficient data ingestion and query processing for general-purpose LSM-based storage systems.
We first propose and evaluate a series of optimizations for efficient batched point lookups,
significantly improving the range of applicability of LSM-based secondary indexes.
We then present several new and efficient maintenance strategies for LSM-based storage systems.
Finally, we have implemented and experimentally evaluated the proposed techniques in the context of the Apache AsterixDB system,
and we present the results here.
\end{abstract}

\section{Introduction}
A wide range of applications, such as risk management, online recommendations, and location-based advertising,
demand the capability of performing real-time analytics on high-speed, continuously
generated data coming from sources such as social networks, mobile devices and IoT applications.
As a result, modern Big Data systems need to efficiently support
both fast data ingestion and real-time queries.

The Log-Structured Merge (LSM) tree~\cite{lsm1996} is a promising structure to support write-intensive workloads.
It has has been widely adopted by NoSQL and NewSQL systems~\cite{cassandra,hbase, leveldb, rocksdb,asterixdb-storage2014,bigtable}
for its superior write performance.
Instead of updating data in-place, which can lead to expensive random I/Os, LSM writes,
including inserts, deletes and updates, are first accumulated in memory
and then subsequently flushed to disk and later merged using sequential I/Os.
A number of improvements have been proposed to optimize various aspects of the original LSM proposal~\cite{compaction2015, monkey2017,dostoevsky2018, blsm2012, lsm-trie2015}.

Auxiliary structures, such as secondary indexes, are critical to enable the efficient processing of ad-hoc queries.
Two types of LSM-based auxiliary structures have been used in practice to facilitate query processing, namely secondary indexes and filters.
A secondary index is an LSM-tree that maps secondary key values to their corresponding primary keys.
A filter, such as a Bloom filter~\cite{bloom-filter1970} or a range filter~\cite{asterixdb-filter2015} on secondary keys, is directly built into LSM-trees to enable data skipping for faster scans.
While needed for queries, maintaining these structures during data ingestion comes with extra cost.
Especially in the case of updates, both types of structures require accessing old records so that they can be properly maintained.
Existing LSM-based systems, such as AsterixDB~\cite{asterixdb-web, asterixdb2014}, MyRocks~\cite{myrocks}, and Phoenix~\cite{phoenix},
employ an eager strategy to maintain auxiliary structures by prefacing each incoming write with a point lookup.
This strategy is straightforward to implement and optimizes for query performance since auxiliary structures are always up-to-date.
However, it leads to significant overhead during ingestion because of the point lookups.

An outage with respect to the general-purpose use of LSM indexing is
that no particularly efficient point lookup algorithms have been proposed for the efficient fetching of the records identified by a secondary index search.
While sorting fetch lists based on primary keys is a well-known optimization~\cite{btree2011},
performing the subsequent point lookups independently still incurs high overhead.
This limits the range of applicability of LSM-based secondary indexes,
requiring the query optimizer to maintain accurate statistics
to make correct decisions for processing ad-hoc queries efficiently.

This paper focuses on efficient data ingestion and query processing techniques for general-purpose LSM-based storage systems.
The first contribution of this paper is to propose and evaluate a series of optimizations
for efficient index-to-index navigation for LSM-based indexes.
We show how to leverage the internal structure of LSM to efficiently process a large number of point lookups.
We further conduct a detailed empirical analysis to evaluate the effectiveness of these optimizations.
The experimental results show that the proposed optimizations greatly improve the range of applicability of LSM-based secondary indexes.
Even with relatively large selectivities, such as 10 - 20\%, LSM-based secondary indexes can still provide performance
better than or (worst case) comparable to that of a full scan.

The second contribution of this paper is to study alternative maintenance strategies for LSM-based auxiliary structures.
The key insight here is the applicability of a \emph{primary key index}, which only stores primary keys, for use in index maintenance.
For secondary indexes, we present a \emph{validation} strategy that cleans up obsolete entries lazily using a primary key index.
For filters, we introduce a \emph{mutable-bitmap} strategy that allows deleted keys to be directly reflected for immutable data through mutable bitmaps
by accessing primary keys instead of full records.
Since primary keys are much smaller than full records,
we show that the exploitation of a primary key index can greatly reduce required I/Os and significantly increase the overall ingestion throughput.

Finally, we have implemented all of the proposed techniques inside Apache AsterixDB~\cite{asterixdb-web, asterixdb2014},
an open-source LSM-based Big Data Management System.
We have conducted extensive experiments to evaluate their impacts on ingestion performance and query performance.

The remainder of the paper is organized as follows:
\refsection{sec:background} provides background information and surveys the related work.
\refsection{sec:architecture} presents a general architecture for LSM-based storage systems
and introduces various optimizations for efficient point lookups.
Sections~\ref{sec:validation} and~\ref{sec:mutable-bitmap} describe in detail
the proposed Validation strategy and the Mutable-bitmap strategy.
\refsection{sec:evaluation} experimentally evaluates the proposed techniques.
Finally, \refsection{sec:conclusion} concludes the paper.

\section{Background}
\label{sec:background}
\subsection{Log-Structured Merge Trees}
The LSM-tree~\cite{lsm1996} is a persistent index structure optimized for write-intensive workloads.
In LSM, writes are first buffered into a memory component and then flushed to disk using sequential I/Os when memory is full.
Each flush operation forms a new disk component.
Once flushed, LSM disk components are immutable.
Modifications (inserts, updates and deletes) are therefore handled by inserting new entries into memory.
An insert or update simply adds a new entry with the same key,
while a delete adds an ``anti-matter'' entry~\cite{asterixdb-storage2014} indicating that a key has been deleted.

A query over LSM data has to reconcile the entries with identical keys from multiple components, as entries from the newer components override those from older components.
As disk components accumulate, query performance tends to degrade since more components must be examined.
To counter this, disk components are gradually merged according to a pre-defined \emph{merge policy}.
In general, two types of merge policies are used in practice~\cite{monkey2017,dostoevsky2018},
both of which organize components into ``levels''.
The leveling merge policy maintains one component per level,
and a component in a higher level will be exponentially larger than that of the next lower one.
In contrast, the tiering merge policy maintains multiple components per level;
these components are formed when a series of lower-level components are merged together using a merge operation.

\subsection{Apache AsterixDB}
\label{sec:asterixdb}
Apache AsterixDB~\cite{asterixdb2014} is a parallel, semistructured Big Data Management System (BDMS)
that aims to support ingesting, storing, indexing, querying and analyzing massive amounts of data efficiently.
Here we briefly discuss storage management in AsterixDB~\cite{asterixdb-storage2014},
which relates to this paper.

The records of a dataset in AsterixDB are hash-partitioned based on their primary keys across multiple nodes.
Each partition of a dataset uses a primary LSM-based \btree index to store the records.
Secondary indexes, including LSM-based \btrees, R-trees, and inverted indexes,
are local to the primary index partition, as in most shared-nothing parallel databases.
The memory components of all the indexes of a partition of a dataset share a memory budget,
and thus they are always flushed together.
A primary index entry is stored as the primary key plus the record, while
a secondary index entry is composed of the secondary key and the associated primary key.
Secondary index lookups are routed to all dataset partitions to fetch the corresponding primary keys.
The returned primary keys are then sorted locally before retrieving the records in the local partitions in order to retrieve them in an efficient manner.

AsterixDB supports record-level, ACID transactions across multiple LSM indexes of a dataset to ensure
that all secondary indexes are consistent with the primary index.
It utilizes a no-steal/no-force buffer management policy with write-ahead-logging (WAL) to ensure durability and atomicity.
For recovery and rollback, index-level logical logging and component shadowing are employed.
Rollback for in-memory component changes is implemented by applying the inverse operations of log records in the reverse order.
To perform crash recovery, the system first examines all valid disk components to compute the maximum component LSN of the index.
Committed transactions beyond the maximum component LSN are then replayed by examining log records.
No undo is performed during recovery since the no-steal policy guarantees that disk components can only contain committed transactions.

\subsection{Related Work}
A number of improvements of the original LSM-tree index structure~\cite{lsm1996} have been proposed recently.
bLSM~\cite{blsm2012} presented a spring-and-gear merge scheduler to reduce periodic write stalls.
cLSM~\cite{clsm2015} is optimized for multi-core machines using non-blocking concurrency control mechanisms.
Monkey~\cite{monkey2017} optimized the memory allocation of Bloom filters for LSM-trees.
Dostoevsky~\cite{dostoevsky2018} presented a lazy-leveling merge policy that can make better performance trade-offs.
WiscKey~\cite{wisckey2017} separated values from keys to reduce write amplification.
All of these efforts have been focused on improving a single LSM index.
Our work here is orthogonal to these optimizations of LSM-trees since we focus on LSM-based auxiliary structures
for efficient query processing.

DELI~\cite{deli2015} presented a lazy secondary index maintenance strategy by repairing secondary indexes during the merge of primary index components.
In addition to its eager strategy, AsterixDB~\cite{asterixdb-storage2014} supports a deleted-key \btree strategy
that attaches a \btree to each secondary index component that records the deleted keys in this component.
Since DELI and the deleted-key \btree strategy are closely related to our work, we will further discuss them in detail in \refsection{sec:validation-overview}.
Qadar et al.~\cite{secondary2018} conducted an experimental study of LSM-based secondary indexes.
However, their study did not consider cleaning up secondary indexes in the case of updates.

Alsubaiee et al.~\cite{asterixdb-filter2015} added the option of a range filter on LSM-based primary and secondary indexes for the efficient processing of time-correlated queries.
Jia~\cite{jia2017} exploited LSM range filters to accelerate primary key lookups for append-only and time-correlated workloads.
However, the eager strategy for maintaining filters incurs high lookup cost during ingestion,
and reduces their pruning capabilities in the presence of updates.
Several commercial database systems
have similarly supported range filter-like structures,
such as zone maps in Oracle~\cite{oracle-zonemap2017}
and synopses in DB2 BLU~\cite{blu2013}, to enable data skipping during scans. However, these systems are not based on LSM-trees.

Several related write-optimized indexes~\cite{chronicle2017, cr-index2014,cr-index2016} have been proposed to efficiently
index append-only observational streams and time-series data.
Though they provide high ingestion performance, updates and deletes are not supported. 
Our work targets general workloads that include updates and deletes as well as appends.

\section{LSM Storage Architecture}
\label{sec:architecture}
In this section, we present a general LSM-based storage architecture that will be the foundation for the rest of this paper.
We also present the Eager strategy used in existing systems as well as a series of optimizations for index-to-index navigation given LSM-based indexes.

\begin{figure}[b]
	\centering
	\includegraphics[width=\linewidth]{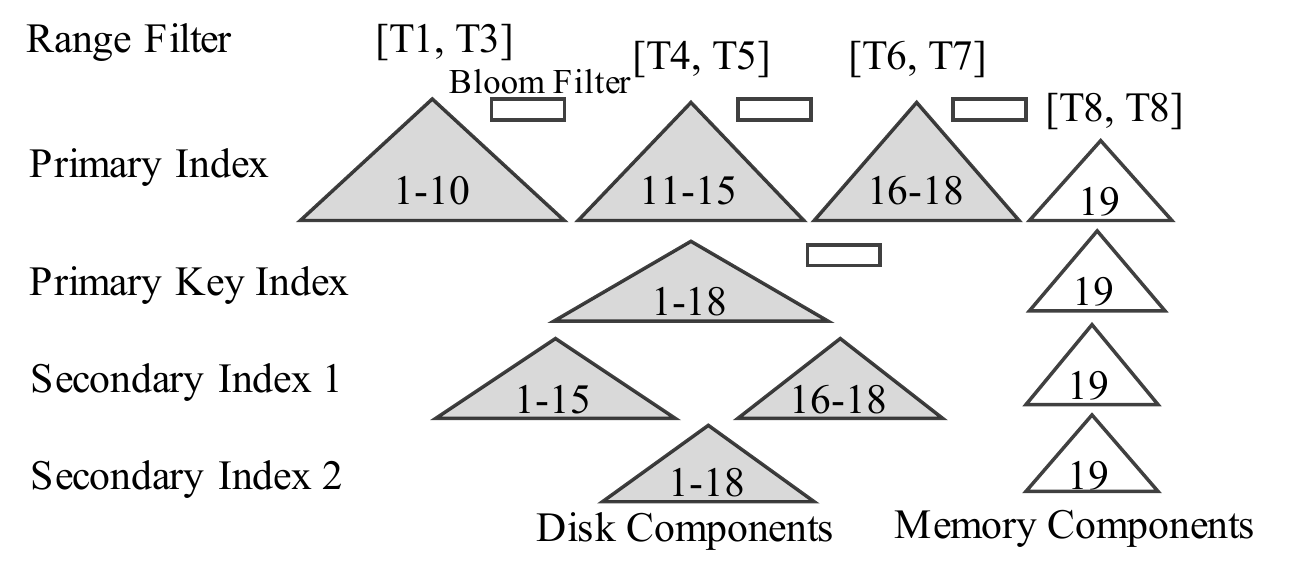}
	\caption{Storage Architecture}
	\label{fig:storage-architecture}
\end{figure}

The storage architecture is depicted in \reffigure{fig:storage-architecture}.
Each dataset has a primary index, a primary key index, and a set of secondary indexes,
which are all based on LSM-trees.
All indexes in a dataset share a common memory budget for memory components, so they are always flushed together (as in AsterixDB).
Each component in the figure is labeled with its \emph{component ID}, which is represented as a pair of timestamps (minTS - maxTS).
The component ID is simply maintained as the minimum and maximum timestamps of the index entries stored in the component,
where the timestamp is generated using the local wall clock time when a record is ingested.
Through component IDs, one can infer the recency ordering among components of different indexes,
which can be useful for secondary index maintenance operations (as discussed later).
For example, component IDs indicate that component 1-15 of Secondary Index 1 is older than component 16-18 of the primary index
and that it overlaps component 1-10 of the primary index.

The primary index stores the records indexed by their primary keys.
To reduce point lookups during data ingestion, as we will see later, we further build a primary key index that stores primary keys only.
Both of these indexes internally use a \btree to organize the data within each component.
Each primary or primary key disk component also has a Bloom filter~\cite{bloom-filter1970} on the stored primary keys to speed up point lookups.
A point lookup query can first check the Bloom filter of a disk component
and search the actual \btree only when the Bloom filter reports that the key may exist.
Secondary indexes use a composition of the secondary key and the primary key as their index key in order to efficiently handle duplicate secondary keys.
A secondary index query can first search the secondary index to return a list of matching primary keys
and then perform point lookups to fetch records from the primary index.

In general, the primary index may have a set of filters for efficient pruning during scans\footnote{As suggested by~\cite{asterixdb-filter2015}, secondary indexes could have filters as well. However, in this paper we only focus on the use of filters on the primary index to support efficient scans.}.
Without loss of generality, we assume that each primary index component may have a range filter that stores the minimum and maximum values of the component's secondary filter key (denoted as [Ti, Tj] in \reffigure{fig:storage-architecture}).
During a scan, a component can be pruned if its filter is disjoint with the search condition of a query.

\subsection{Data Ingestion with the Eager Strategy}
\label{sec:eager}
During data ingestion, all storage structures must be properly maintained.
Here we briefly review the Eager strategy commonly used by existing LSM-based systems such as AsterixDB~\cite{asterixdb-web, asterixdb2014}, MyRocks~\cite{myrocks}, and Phoenix~\cite{phoenix}.

To \emph{insert} a record, its key uniqueness is first checked by performing a point lookup.
As an optimization, the primary key index can be searched instead for efficiency.
If the given primary key already exists, the record is ignored;
otherwise, the record is recorded in the memory components of all of the dataset's indexes.
Any filters of the memory components must be maintained based on the new record as well.

To \emph{delete} a record given its key, a point lookup is first performed to fetch the record.
If the record does not exist, the key is simply ignored.
Otherwise, anti-matter entries must be inserted into all of the dataset's LSM indexes to delete the record.
Any filters of the memory components must also be maintained based on the deleted record;
otherwise, a query could erroneously prune a memory component and thus access deleted records.

To \emph{upsert} a record, a point lookup is first performed to locate the old record with the same key.
If the old record exists, anti-matter entires are generated to delete the old record from all of the dataset's secondary indexes.
The new record is then inserted into the memory components of all of the dataset's LSM indexes.
As an optimization, if the value of some secondary key did not change,
the corresponding secondary index can be simply skipped for maintenance.
Any filters of the memory components must be maintained based
on both the old record (if it exists) and the new record.

\begin{figure}[t]
	\begin{minipage}{\linewidth}
		\centering
		\includegraphics[width=0.85\linewidth]{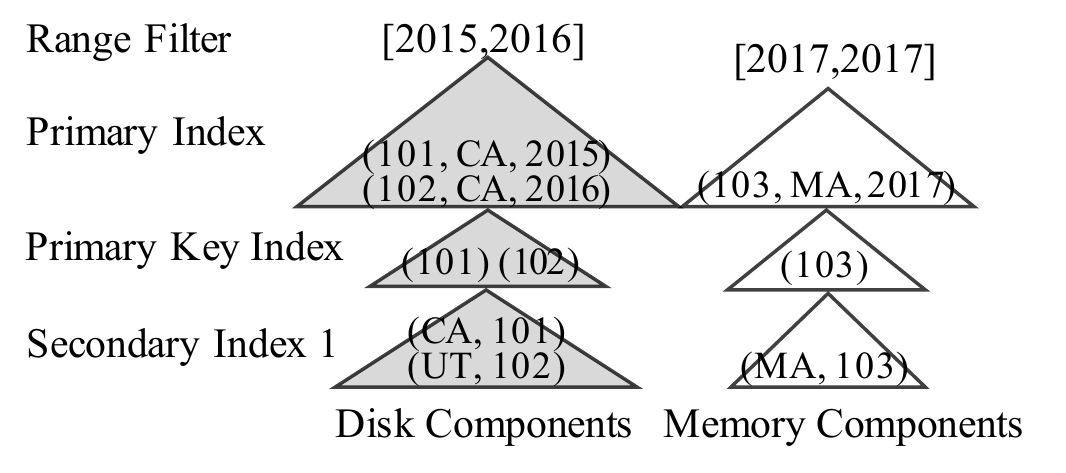}
		\caption{Running Example}
		\label{fig:example-original}
	\end{minipage}
	\begin{minipage}{\linewidth}
		\centering
		\includegraphics[width=0.85\linewidth]{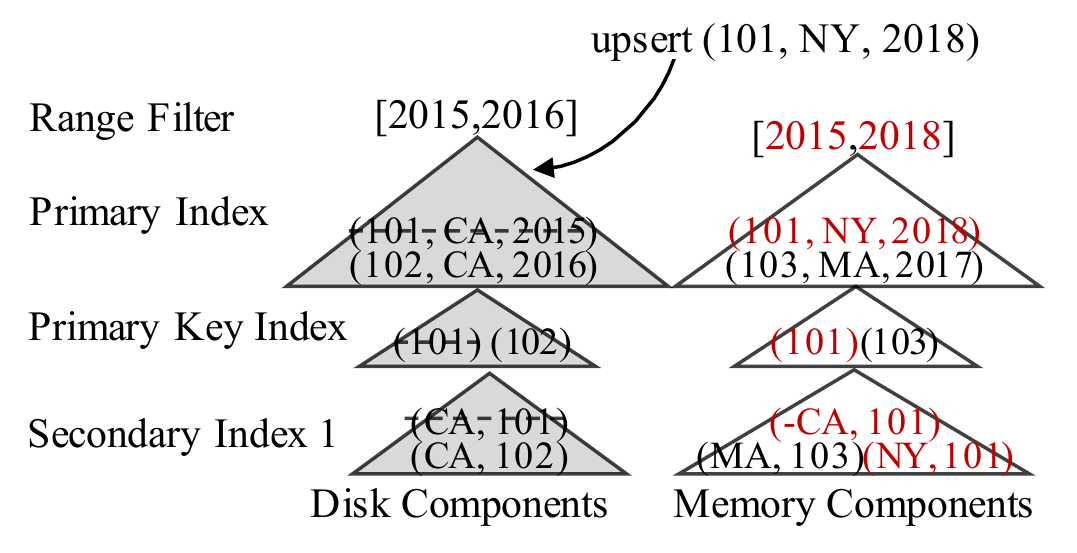}
		\caption{Upsert Example with Eager Strategy}
		\label{fig:example-eager}
	\end{minipage}
\end{figure}

As a running example, consider the initial LSM indexes depicted in \reffigure{fig:example-original} for a UserLocation dataset
with three attributes: UserID (the primary key), Location, and Time.
The Location attribute stores the last known location of the user (in terms of states) and the Time attribute stores the time of the last known location (in terms of years).
We have a secondary index on Location and a range filter on Time.
\reffigure{fig:example-eager} shows the resulting LSM indexes after upserting a new record (101, NY, 2018) with an existing key 101.
A point lookup is first performed to locate the old record (101, CA, 2015).
In addition to adding the new record to all memory components, an anti-matter entry (-CA, 101) is added to the secondary index to eliminate the obsolete entry (which is dashed in the figure).
Also, notice that the memory component's range filter is maintained based on both 2015 and  2018
so that future queries will properly see that the old record with UserID 101 has been deleted.

We further use \reffigure{fig:example-eager}'s example to illustrate how secondary indexes and filters are used in query processing.
First consider a query Q1 that finds all user records with Location CA.
Q1 first searches the secondary index to return a list of UserIDs.
In this example only UserID 102 is returned since (CA, 101) is deleted by the anti-matter entry (-CA, 101).
The primary index is then searched to fetch the record using UserID 102, which returns (102, CA, 2016).
Note that without the anti-matter entry, the obsolete UserID 101 would be erroneously searched by Q1 as well.
Consider another query Q2 that finds all records with Time $<$ 2017.
Q2 scans the primary index by first collecting a set of candidate components whose range filters overlap with the search condition Time $<$ 2017.
In this example, both components will be scanned by Q2 and only one record (102, CA, 2016) is returned.
However, suppose that when upserting record (101, NY, 2018), the range filter of the memory component had only been maintained based on the new value 2018.
In this case, the memory component would have been pruned and (101, CA, 2015) would be erroneously returned as well.

\subsection{Efficient Index-to-index Navigation}
\label{sec:index-navigation}
Navigating from secondary indexes to the primary index is a fundamental operation for query processing.
Traditionally, primary keys are sorted to ensure that the pages of the primary index will be accessed sequentially~\cite{btree2011}.
Here we discuss further optimizations to improve point lookup performance for LSM-trees.
Note that some of the optimizations below are not new.
Our contribution is to evaluate their effectiveness and integrate them to improve the range of applicability of LSM-based secondary indexes.

\textbf{Batched Point Lookup.}
Even though primary keys are sorted, when searching multiple LSM components, it is still possible that index pages will be fetched via
random I/Os since the sorted keys can be scattered across different components.
To avoid this, we propose here a \emph{batched point lookup} algorithm that works as follows:
Sorted primary keys are first divided into batches.
For each batch, all of the LSM components are accessed one by one, from newest to oldest.
Specifically, for each key in the current batch that has not been found yet, 
it is searched against a primary component by first checking the Bloom filter and then the \btree.
A given batch terminates either when all components have been searched or all keys have been found.
The batch search algorithm ensures that components' pages are accessed sequentially,
avoiding random I/Os when fetching their leaf pages.
However, a downside is that the returned data records will no longer still be ordered on primary keys.
We will experimentally evaluate this trade-off in \refsection{sec:evaluation}.

\textbf{Stateful \btree Lookup.}
To reduce the in-memory \btree search overhead, one can use a stateful \btree search cursor that remembers the search history from root to leaf.
Instead of always traversing from the root for each key, the cursor starts from the last leaf page to reduce the tree traversal cost.
One can further use exponential search~\cite{exponential-search1976} instead of binary search to reduce the search cost within each page.
To search a key, this algorithm starts from the last search position and searches for the key using exponentially increasing steps
to locate the range which this key resides in.
The key can then be located by performing a binary search within this range.

\textbf{Blocked Bloom Filter.}
Finally, to reduce the overhead of checking the components' Bloom filters,
a cache-friendly approach called blocked Bloom filter~\cite{block-bloomfilter2009} can be used.
The basic idea is to divide the bit space into fixed-length blocks
whose size is the same as the CPU cache line size.
The first hash function maps a key to a block, while the rest of the hash functions perform the usual bit tests but within this block.
This ensures that each Bloom filter test will only lead to one cache miss, at the cost of requiring an extra bit per key to achieve the same false positive rate~\cite{block-bloomfilter2009}.

\section{Validation Strategy}
\label{sec:validation}
In this section, we propose the Validation strategy for maintaining secondary indexes efficiently.
We first present an overview followed by detailed discussions of this strategy.

\subsection{Overview}
\label{sec:validation-overview}
In a primary LSM-tree, an update can blindly place a new entry (with the identical key) into memory to mark the old entry as obsolete.
However, this mechanism does not work with secondary indexes, as the index key could change after an update.
Similarly, one cannot efficiently determine whether a given primary key is still valid based on a secondary index alone,
as entries in a secondary index are ordered based on secondary keys.
Extra work must be performed to maintain secondary indexes during data ingestion.

The Eager strategy maintains secondary indexes by producing anti-matter entries for each old record,
which incurs a large point lookup overhead during data ingestion.
An alternative strategy is to only insert new entries into secondary indexes
during data ingestion while cleanup obsolete entries lazily so that the expensive point lookups can be avoided.
This design further ensures that secondary indexes can only return false positives (obsolete primary keys) but not false negatives, which simplifies query processing.
Although the idea of lazy maintenance is straightforward, two challenges must be addressed:
(1) how to support queries efficiently, including both non-index-only and index-only queries;
(2) how to repair secondary indexes efficiently to cleanup obsolete entries while avoiding making cleanup a new bottleneck.

There have been several proposals for lazy secondary index maintenance on LSM-based storage systems.
DELI~\cite{deli2015} maintains secondary indexes lazily, while merging the primary index components, without introducing any additional structures.
If multiple records with the same primary key are encountered during a merge, DELI produces anti-matter entries for obsolete records to clean up secondary indexes.
However, this design does not completely address the above two challenges.
First, non-index-only queries cannot be supported efficiently, as queries must fetch records to validate the search results.
Second, DELI lacks the flexibility to repair secondary indexes efficiently.
For update-heavy workload, it is desirable to repair secondary indexes more frequently to improve query performance.
However, in DELI this requires constantly merging or scanning all components of the primary index, incurring a high I/O cost.

To support index-only queries, extra structures on top of secondary indexes should be maintained.
AsterixDB supports a deleted-key \btree strategy
that attaches a \btree to each secondary index component that records the deleted keys in this component.
For index-only queries, validation can be performed by searching these deleted-key \btrees without accessing full records.
However, since these \btrees are duplicated for each secondary index, and no efficient repairing algorithms have been described,
cleaning up secondary indexes incurs a high overhead during merges.

To address the aforementioned challenges,
we choose to use the primary key index to maintain all secondary indexes efficiently.
The primary key index can be used for validating index-only queries by storing an extra timestamp
for each index entry in all secondary indexes as well as in the primary key index.
This timestamp is generated using the node-local clock time when a record is ingested and
is stored as an integer with 8 bytes.
We further propose an efficient index repair algorithm based on the primary key index, greatly reducing the I/O cost by avoiding accessing full records.

\subsection{Data Ingestion}
Under the Validation strategy, an \emph{insert} is handled exactly as in the Eager strategy
except that timestamps are added to the primary key index entries and secondary index entries.
To \emph{delete} a record given its key, an anti-matter entry is simply inserted into both the primary index and the primary key index.
To \emph{upsert} a record, the new record is simply inserted into all of the dataset's LSM indexes.
(Notice that when deleting or upserting a record, the memory component of the primary index has to be searched to find the location for the entry being added.
As an optimization, then, if the old record happens to reside in the memory component, 
it can be used to produce local anti-matter entries to clean up the secondary indexes without additional cost.)

\begin{figure}
	\centering
	\includegraphics[width=0.85\linewidth]{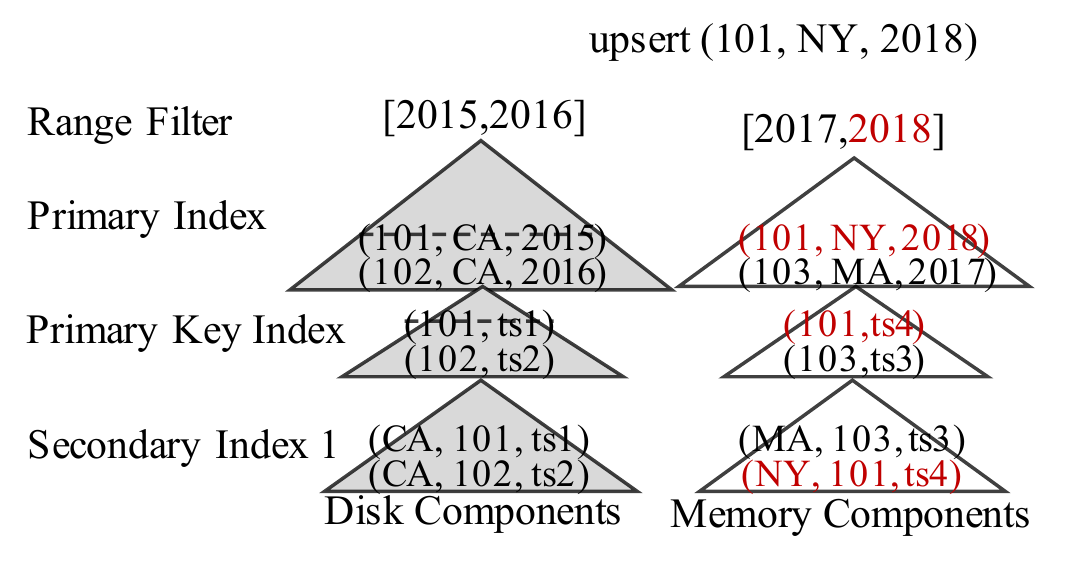}
	\caption{Upsert Example with Validation Strategy}
	\label{fig:example-validation}
\end{figure}

Consider the running example that began in \reffigure{fig:example-original}.
\reffigure{fig:example-validation} shows the resulting LSM indexes after upserting the record (101, NY, 2018) under the Validation strategy.
The primary key index and the secondary index each now contain an extra timestamp field (denoted as ``ts'').
To upsert the new record, we add the new record to all memory components without any point lookups.
As a result, the obsolete secondary index entry (CA, 101, ts1) still appears to be valid even though it points to a deleted record.
The Validation strategy can be naturally extended to support filters:
Since no pre-operation point lookup is performed, the filters of the memory components are only maintained based on new records.
As in the example, then, the range filter is only maintained based on 2018.
As a result, a query that accesses an older component has to access all newer components in order not to miss any newer overriding updates.

\subsection{Query Processing}
\label{sec:validation-query}
\begin{figure}
	\centering
	\begin{subfigure}{.24\textwidth}
		\centering
		\includegraphics[width=0.72\linewidth]{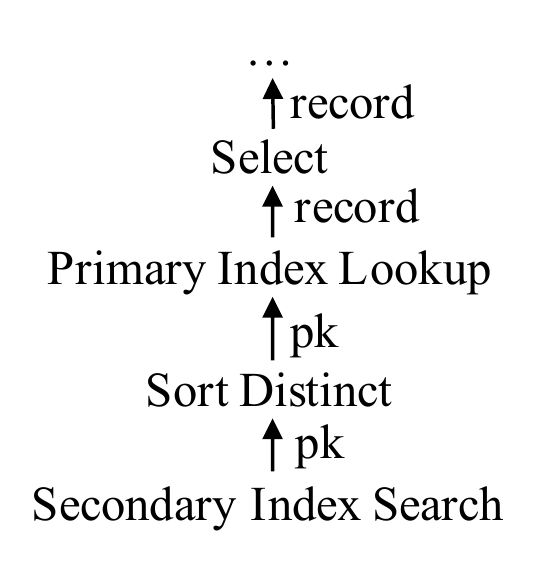}
		\caption{Direct Validation}
		\label{fig:direct-validation}
	\end{subfigure}%
	\begin{subfigure}{.24\textwidth}
		\centering
		\includegraphics[width=0.9\linewidth]{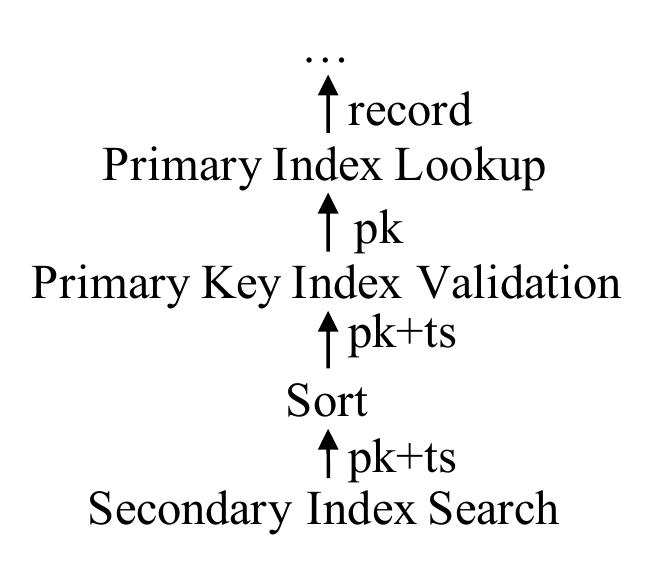}
		\caption{Timestamp Validation}
		\label{fig:ts-validation}
	\end{subfigure}
	\caption{Query Validation Methods}
	\label{fig:query-validation}
\end{figure}

In the Validation strategy, secondary indexes are not always guaranteed to be up-to-date and can thus return obsolete entries to queries.
For correctness, queries have to perform an extra validation step to ensure that only valid keys are eventually accessed.
Here we present two validation method variations suitable for different queries (\reffigure{fig:query-validation}).

The Direct Validation method (\reffigure{fig:direct-validation}) 
directly performs point lookups to fetch all of the candidate records and re-checks the search condition.
A sort-distinct step is performed first to remove any duplicate primary keys.
After checking the search condition, only the valid records are returned to the query.
However,  this method rules out the possibility of supporting index-only queries efficiently.

To address these drawbacks, the Timestamp Validation method (\reffigure{fig:ts-validation}) uses the primary key index to perform validation.
A secondary index search returns the primary keys plus their timestamps.
Point lookups against the primary key index are then performed to validate the fetched primary keys.
Specifically, a key is invalid if the same key exists in the primary key index but with a larger timestamp.
The valid keys are then used to fetch records if necessary.

Consider the example in \reffigure{fig:example-validation},
and a query that wants to find all records with Location CA.
The secondary index search returns primary keys with their timestamps (101, ts1) and (102, ts2).
Direct Validation performs point lookups to locate records (101, NY, 2018) and (102, CA, 2016).
The first record will be filtered out because its Location is not CA anymore.
Timestamp Validation would perform a point lookup against the primary key index to filter out UserID 101
since its timestamp ts1 is older than ts4.

\subsection{Secondary Index Repair}
\label{sec:index-repair}
Since secondary indexes are not cleaned up during ingestion under the Validation strategy,
obsolete entries could accumulate and degrade query performance.
To address this, we propose performing index repair operations in the background to clean up obsolete index entries.
Index repair can either be performed during merge time, which we will call \emph{merge repair},
or scheduled independently from merges, which we will call \emph{standalone repair}.

\begin{figure}
		\centering
		\includegraphics[width=\linewidth]{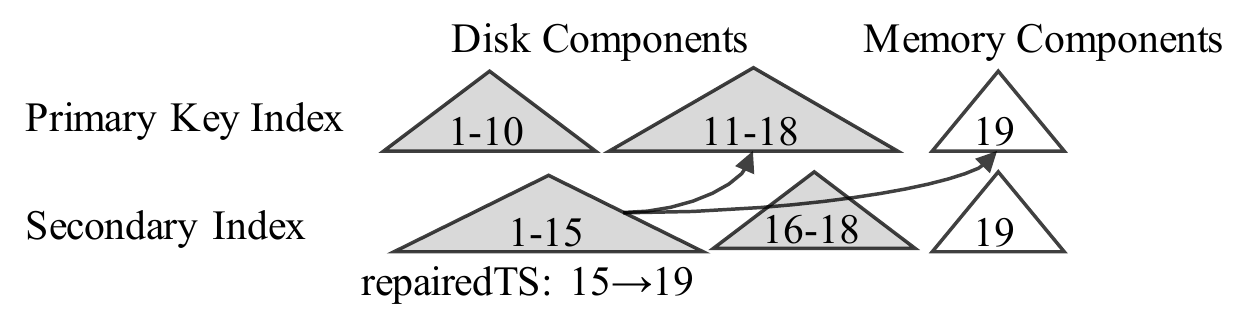}
		\caption{Repaired Timestamp Example}
		\label{fig:repaired-ts}
\end{figure}

The basic idea of index repair is to validate each primary key in a component by searching the primary key index.
For efficiency, the primary keys in a given component of a secondary index should only be validated against newly ingested keys.
To keep track of the repair progress, we associate a \emph{repaired timestamp} (repairedTS) with each disk component of a secondary index.
During a repair operation, all primary key index components with maxTS no larger than the repairedTS can be pruned.\footnote{This optimization applies to Timestamp Validation as well.}
A repaired component receives a new repairedTS computed as the maximum timestamp of the unpruned primary key index components.
To clarify this, consider the example in \reffigure{fig:repaired-ts}.
Component 1-15 of the secondary index has an initial repairedTS of 15.
To repair this component, we only need to search components 11-18 and 19 of the primary key index, while component 1-10 can be pruned.
After the repair operation, the new repairedTS of this component would be 19.

A naive implementation of merge repair would simply validate each primary key by performing a point lookup against the primary key index.
For standalone repair, one could simply produce a new component with only valid entries without merging.
However, this implementation would be highly inefficient because of the expensive random point lookups.

To handle this, we propose a more efficient merge repair algorithm (\reffigure{alg:merge-repair}).
The basic idea is to first sort the primary keys to improve point lookup performance,
and further use an immutable bitmap to avoid having to sort entries back into secondary key order.
The immutable bitmap of a disk component indicates whether each of the component's index entries is (still) valid or not,
and thus it stores one bit per entry.
Without loss of generality, we assume that a bit being 1 indicates that the corresponding index entry is invalid.

\begin{figure}
	\centering
	\small
	\begin{algorithmic}[1]
		\State{Create search cursor on merging components}
		\State{position $\gets$ 0}
		\While{cursor.hasNext()}
		\State{entry $\gets$ cursor.getNextEntry()}
		\State{add entry to new component}
		\State{add (pkey, ts, position) to sorter}
		\State{position $\gets$ position + 1}
		\EndWhile
		\State{initialize bitmap with all 0s}
		\State{sorter.sort()}
		\For{sorted entries (pkey, ts, position)}
		\State{validate pkey against primary key index}
		\If{pkey is invalid}
		\State{mark position of bitmap to 1}
		\EndIf
		\EndFor
	\end{algorithmic}
	\caption{Pseudo Code for Merge Repair}
	\label{alg:merge-repair}
\end{figure}

Initially, we create a scan cursor over all merging components to obtain all valid index entries,
i.e., entries where their immutable bitmap bits are 0.
These entries are directly added to the new component (line 5).
Meanwhile, the primary keys, with their timestamps and positions in the new component,
are streamed to a sorter (line 6).
These sorted primary keys are then validated by searching the primary key index.
As an optimization, if the number of primary keys to be validated is larger than the number of recently ingested keys
in the primary key index, we can simply merge scan the sorted primary keys and the primary key index.
If a key is found to be invalid, that is, if the same key exists in the primary key index with a larger timestamp,
we simply set the corresponding position of the new component's bitmap to 1 (lines 12-13).
Standalone repair can be implemented similarly, except that only a new bitmap is created.

\begin{figure}
		\centering
		\includegraphics[width=\linewidth]{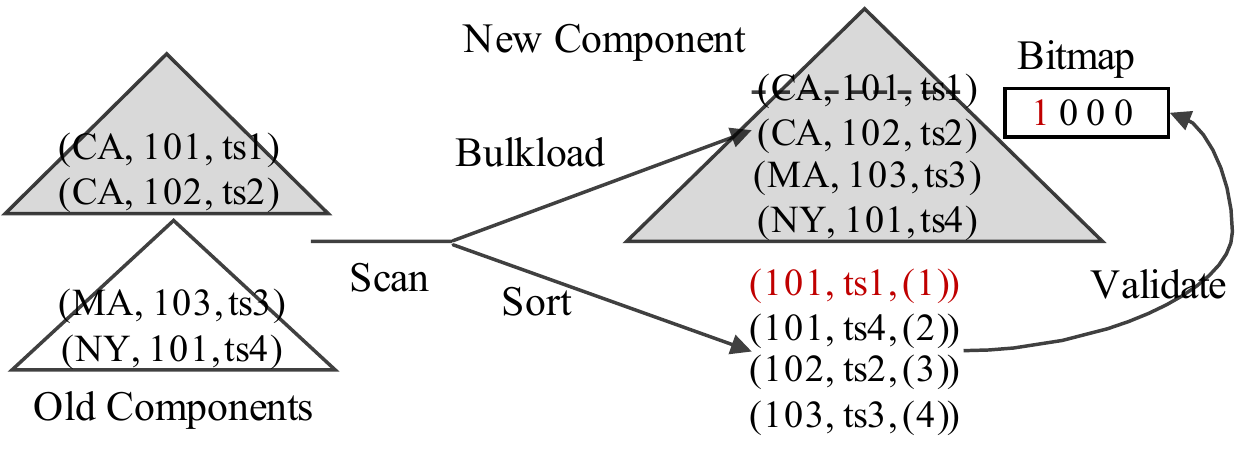}
		\caption{Merge Repair Example}
		\label{fig:repair-example}
\end{figure}

To illustrate the immutable bitmap and the repair process, consider the example secondary index in \reffigure{fig:example-validation}.
Suppose we want to merge and repair all components of the secondary index.
The process is shown in \reffigure{fig:repair-example}.
The index entries scanned from its old components are directly added to the new component,
and in the meanwhile they are sorted into primary key order.
The sorted primary keys are validated using the primary key index.
During validation, the key 101 with timestamp ts1 is found to be invalid since the same key exists with a larger timestamp ts4.
This index entry has ordinal position 1 (denoted ``(1)'' in the figure), so first bit in the component's bitmap is set to 1.
Note that this invalid entry will be physically removed during the next merge.

\textbf{Bloom Filter Optimization.}
It is tempting to use Bloom filters to further optimize the index repair operation.
The idea is that if the Bloom filters of the primary key index components
do not contain a key, which implies that the key has not been updated,
then the key can be excluded from sorting and further validation.
However, if implemented directly, this would provide little help since a dataset's various LSM-trees are merged independently.
Consider the example LSM indexes back in \reffigure{fig:storage-architecture}.
Suppose we want to merge and repair the two disk components of Secondary Index 1.
Since the disk components of the primary key index have already been merged into a single component beforehand,
its Bloom filter would always report positives, which would provide no help and actually cause some extra overhead.

To maximize the effectiveness of the Bloom filter optimization, one must ensure that
during each repair operation the unpruned primary key index components are always strictly newer than the keys in the repairing component(s).
For this purpose, we can use a correlated merge policy~\cite{asterixdb-filter2015} to 
synchronize the merge of all secondary indexes with the primary key index in order to ensure that their components are always merged together.
Furthermore, all secondary indexes must be repaired during every merge.
As a side-effect of this optimization, the timestamps of index entries can be discarded since validations can be performed by using the relative ordering among components.
For example, in \reffigure{fig:example-validation} the secondary index entry (v1, k1) is invalid since the same key k1 exists in a newer (memory) component of the primary key index.

The Bloom filter optimization improves repair efficiency and is thus suitable for update-heavy workloads that require secondary indexes to be frequently repaired.
However, for workloads that contain few updates, it might be better to just schedule repair operations during off-peak hours,
and thus the Bloom filter optimization may not be suitable.
To alleviate the tuning effort required from the end-user, we also plan to develop auto-tuning techniques in the future.

\section{Mutable-Bitmap Strategy}
\label{sec:mutable-bitmap}
In this section, we present the Mutable-bitmap strategy designed for maintaining a primary index with filters.

\subsection{Overview}
The key difficulty of applying filters to the LSM-tree is its out-of-place update nature, that is, updates are added to the new component.
In the case of updates, the filter of the new component must be maintained using old records so that queries would not miss any new updates.
The Eager strategy performs point lookups to maintain filters using old records,
incurring a high point lookup cost during data ingestion.
The Validation strategy skips point lookups but requires queries to access all newer components for validation,
halving the pruning capabilities of filters.

The Mutable-bitmap strategy presented below aims at both maximizing the pruning capabilities of filters and
reducing the point lookup cost during data ingestion.
The first goal can be achieved if old records from disk components can be deleted directly.
However, if we were to place new updates directly into the disk components where old records are stored,
a lot of complexity would be introduced on concurrency control and recovery.
Instead, our solution is to add a mutable bitmap to each disk component to indicate the validity of each entry using a very limited degree of mutability.
We present efficient solutions to address concurrency control and recovery issues, exploiting the simple semantics of mutable bitmaps, that is, writers only change bits from 0 to 1 to mark records as deleted\footnote{Aborts internally change bits from 1 to 0.}.
To minimize the point lookup cost,
the maintenance of mutable bitmaps is performed by searching the primary key index instead of accessing full records.
To achieve this, we synchronize the merges of the primary index and the primary key index using the correlated merge policy (as in \refsection{sec:index-repair}).
An alternative implementation is to add a key-only \btree to each primary index component, combining two indexes together.

\subsection{Data Ingestion}
For ease of discussion, let us first assume that there are no concurrent flush and merge operations,
the handling of which are postponed to \refsection{sec:concurrent-flush-merge}.
An \emph{insert} is handled exactly as in the Eager/Validation strategy,
and no bitmaps are updated because no record is deleted.
To \emph{delete} a record given its key, we first search the primary key index to locate the position of the deleted key.
If the key is found and is in a disk component, then its corresponding bit in that component's bitmap is set (mutated) to 1.
An anti-matter key is also added to the memory component, for two reasons.
First, we view the mutable bitmap as an auxiliary structure built on top of LSM that should not change the semantics of LSM itself.
Second, if the Validation strategy is used for secondary indexes,
inserting anti-matter entries ensures that validation can be performed with only recently ingested keys.
To \emph{upsert} a record, the primary key index is first searched to mark the old record as deleted if necessary,
and the new record is added to the index memory components.
Any filters of the memory components are only maintained based on the new record, not the old one.

\begin{figure}
	\centering
	\includegraphics[width=0.85\linewidth]{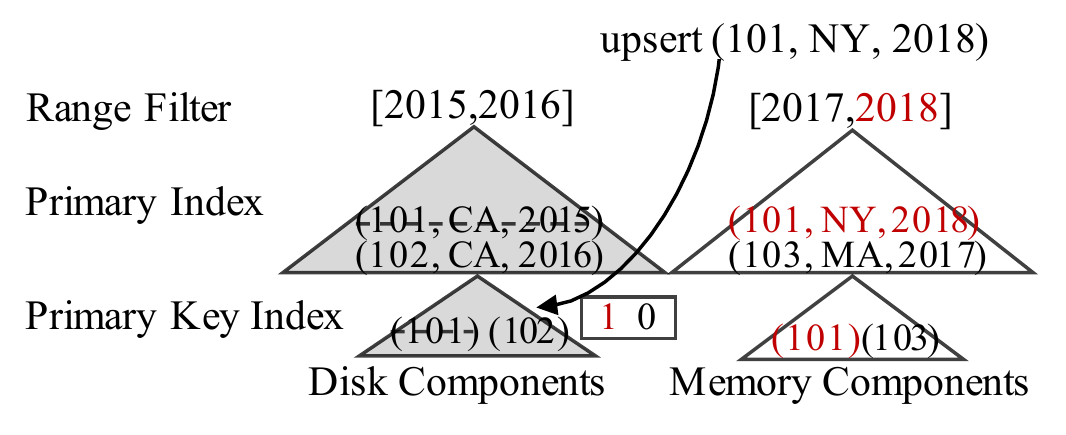}
	\caption{Upsert Example with Mutable-bitmap Strategy}
	\label{fig:example-mutable-bitmap}
\end{figure}

Consider the running example from \reffigure{fig:example-original}.
\reffigure{fig:example-mutable-bitmap} shows the resulting LSM-trees after upserting a new record (101, NY, 2018).
The secondary index is not shown since it can be maintained using either the Eager or Validation strategy.
In this case, the primary index and the primary key index are synchronized and their components share a mutable bitmap to indicate the validity of their records.
To upsert the new record, the primary key index is searched to locate the position of the old record and the bitmap is mutated to mark the old record as deleted.
The range filter of the memory component is maintained based only on the new record, avoiding unnecessary widening based on  2015.

We now discuss concurrency control and recovery issues for the mutable bitmaps.
We assume that each writer acquires an exclusive (X) lock on a primary key throughout the (record-level) transaction.
To prevent two writers from modifying the same bitmap byte, one can use latching or compare-and-swap instructions.
For recovery, we use an additional update bit in the log record for each delete or upsert operation
to indicate whether the key existed in a disk component.
To abort a transaction, if the update bit is 1, we simply perform a primary key index lookup (without bitmaps) to unset the bit from 1 to 0.
To unify the recovery of bitmaps with LSM-trees, we use a no-steal and no-force policy for bitmaps as well.
A modified bitmap page is pinned until the transaction terminates to prevent dirty pages from being flushed.
Regular checkpointing can be performed to flush dirty pages of bitmaps.
Upon recovery, committed transactions are simply replayed to bring bitmaps up-to-date based on the last checkpointed LSN.
Again, a log record is replayed on the bitmaps only when its update bit is 1.

\subsection{Concurrency Control for Flush/Merge}
\label{sec:concurrent-flush-merge}
Mutable bitmaps also introduce concurrency control issues for LSM flush and merge operations.
This is because concurrent writers may need to modify the bitmaps of the components that are being formed by flush or merge operations.
This problem bears some similarities with previous work on online index construction~\cite{online-index1992,perf-online-index1992},
which builds new indexes concurrently with updates.
Here we propose two possible concurrency control methods with different flavors in terms of how new updates are applied to the new component.
The \emph{Lock} method directly applies new updates to the new component at the cost of extra locking overhead. The \emph{Side-file} method buffers new updates in a side-file and then applies them after the new component has been fully built.

\begin{figure}
	\centering
	\small
	\begin{subfigure}[b]{1\linewidth}
		\centering
		\begin{algorithmic}[1]
			\Function{Build}{new component}
			\State{set old component(s) point to new component $C'$}
			\State{create cursor on old component(s)}
			\While{cursor.hasNext()}
			\State{(key, record) $\gets$ cursor.getNext()}
			\State{S lock key}
			\If {key is still valid by checking bitmap}
			\State {add (key, record) to new component}
			\EndIf
			\State{$C'$.ScannedKey $\gets$ key}
			\State{unlock key}
			\EndWhile
			\EndFunction
		\end{algorithmic}
		\caption{Pseudo Code for Component Builder}
		\label{alg:lock-component-builder}
	\end{subfigure} 
	\hfil
	\begin{subfigure}[b]{\linewidth}
		\centering
		\begin{algorithmic}[1]
			\Function{Delete}{key}
			\State{X lock on key}
			\State{search key from the primary key index}
			\If{key exists in immutable component C}
			\State{mark key deleted in C}
				\If{C points to $C'$ AND key $\le$ $C'$.ScannedKey}
				\State{mark key deleted in $C'$}
			\EndIf
			\EndIf
			\State{unlock key}
			\EndFunction
		\end{algorithmic}
		\caption{Pseudo Code for Writer}
		\label{alg:lock-writer}
	\end{subfigure}
	\caption{Pseudo Code for Lock Method}
	\label{fig:lock-method}
\end{figure}

\textbf{Lock Method.} The pseudo code for the Lock method is shown in \reffigure{fig:lock-method},
where the component builder is responsible for building the new component during a flush or merge operation.
The component builder first sets the old component(s) to point to the new component (line 2) so that the new component is visible to writers.
A full scan over the old component(s) is then performed to build the new component.
For each scanned key, the component builder acquires a shared lock (line 6) to prevent the key from being deleted by writers.
Otherwise, if a transaction were to, the deleted key may have already been skipped by the component builder,
making it impossible to rollback in the middle of the bulkloading process.
After the lock has been acquired, the bitmap is re-checked (line 7) to ensure that the key is still valid.
To delete a key, the writer first searches the primary key index.
If the key is found in a disk component $C$, it is marked as deleted (lines 4-5).
If $C$ further points to a new component $C'$ and the deleted key has already been added to $C'$,
the key is further marked as deleted in $C'$ by performing another point lookup (lines 6-7).

\textbf{Side-file Method.}
The pseudo code for the Side-file method is shown in \reffigure{fig:snapshot-method}.
For ease of exposition, we divide the component building process into three phases.
During the initialization phase (lines 2-5), the component builder acquires a shared lock on the dataset to drain ongoing transactions (explained later)
and then creates immutable bitmap snapshots of old components.
During the build phase (lines 6-10), the old components are scanned with the bitmap snapshots 
to avoid interference from concurrent updates.
Finally, during the catchup phase (lines 11-16), a shared lock on the dataset is acquired again to close the side-file.
The component builder sorts the side-file as suggested in~\cite{perf-online-index1992} and then applies the deleted keys to the new component.

Shared locks on the dataset are acquired to ensure correctness in the case of transaction rollbacks.
Otherwise, if a transaction were to abort afterwards, it may not be able to undo its changes to the new components.
If a transaction deletes a key before the initialization phase, but aborts after the bitmap snapshot has been created, 
the deleted key may have already been skipped by the component builder and cannot be re-added to the new component.
Similarly, if a transaction deletes a key during the build phase, but aborts after the side-file has been closed,
the component builder may re-delete this key based on the side-file.

The writer, which deletes a key, is similar to that of the lock method,
except that the deleted key is first appended to the side-file (line 7).
If this fails, which implies that the side-file has been closed, then the deleted key is applied to the new component directly.
In the case of rollback, the transaction simply appends an anti-matter key to the side-file if the side-file is still open.
Otherwise, the transaction simply unsets the bitmap of the new component to 0.

\begin{figure}
	\centering
	\small
	\begin{subfigure}[b]{1\linewidth}
		\centering
		\begin{algorithmic}[1]
			\Function{Initialize}{new component}
			\State{S lock dataset}
			\State{create bitmap snapshots of old component(s)}
			\State{set old component(s) point to new component $C'$}
			\State{unlock dataset}
			\EndFunction{}
			
			\Function{Build}{new component}
			\State{create cursor on old component(s) with bitmap snapshots}
			\While{cursor.hasNext()}
			\State{(key, record) $\gets$ cursor.getNext()}
			\State {add (key, record) to new component $C'$}
			\EndWhile
			\EndFunction
			
			\Function{CatchUp}{new component}
			\State{S lock dataset}
			\State{mark side-file closed}
			\State{unlock dataset}
			\State{sort side-file}
			\State{apply updates in side-file to new component $C'$}
			\EndFunction
		\end{algorithmic}
		\caption{Pseudo Code for Component Builder}
		\label{alg:snapshot-component-builder}
	\end{subfigure} 
	\hfil
	\begin{subfigure}[b]{\linewidth}
		\centering
		\begin{algorithmic}[1]
			\Function{Delete}{key}
			\State{X lock key}
			\State{search key from the primary key index}
			\If{key exists in immutable component $C$}
				\State{mark key deleted in $C$}
				\If{$C$ points to $C'$}
					\State{try to append key to side-file}
					\If{append fails}
						\State{mark key deleted in $C'$}
					\EndIf
				\EndIf
			\EndIf
			\State{unlock key}
			\EndFunction
		\end{algorithmic}
		\caption{Pseudo Code for Writer}
		\label{alg:snapshot-writer}
	\end{subfigure}
	\caption{Pseudo for Side-file Method}
	\label{fig:snapshot-method}
\end{figure}

\begin{figure*}[t]
	\centering
	\begin{subfigure}[b]{.287\linewidth}
		\centering
		\includegraphics[width=\linewidth]{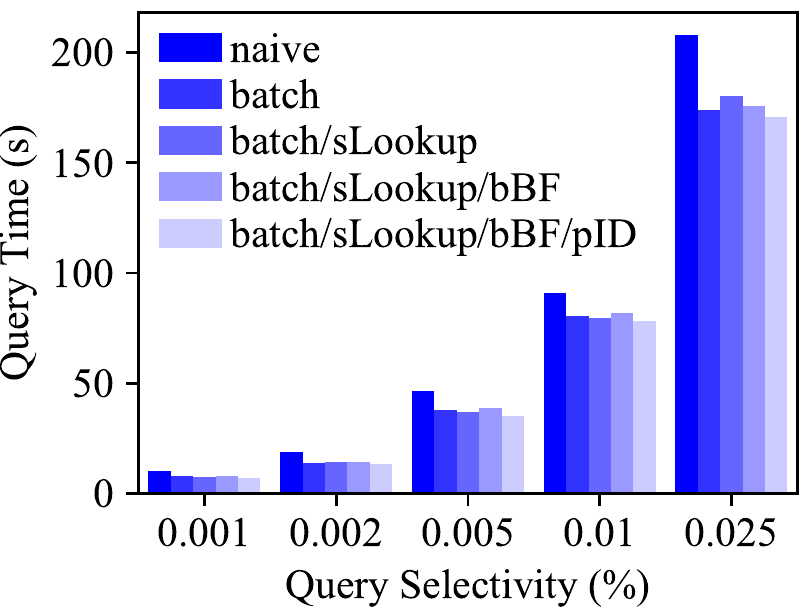}
		\caption{Low Selectivity}
		\label{fig:query-lookup-opt-low}
	\end{subfigure}%
	\begin{subfigure}[b]{.276\linewidth}
		\centering
		\includegraphics[width=\linewidth]{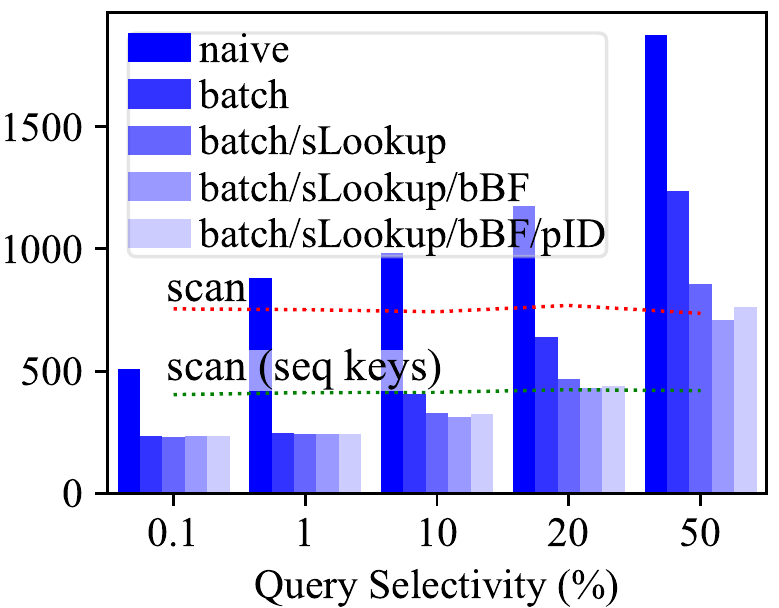}
		\caption{High Selectivity}
		\label{fig:query-lookup-opt-high}
	\end{subfigure}%
	\begin{subfigure}[b]{.223\linewidth}
		\centering
		\includegraphics[width=\linewidth]{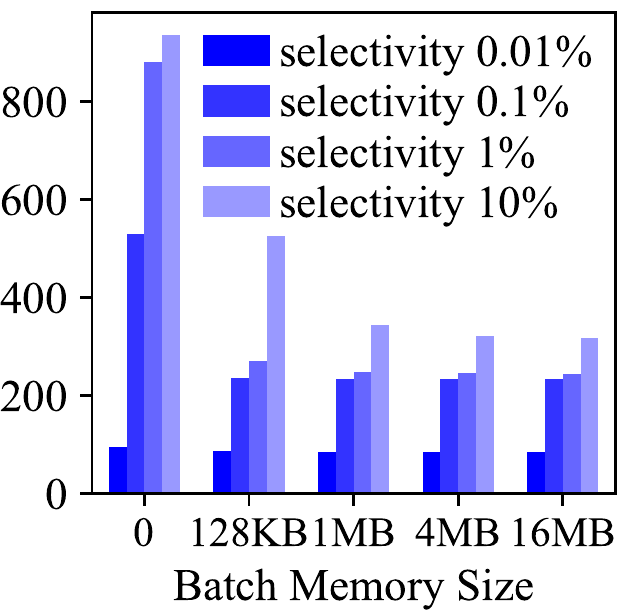}
		\caption{Impact of Batch Size}
		\label{fig:query-batch-size}
	\end{subfigure}%
	\begin{subfigure}[b]{.2083\linewidth}
		\centering
		\includegraphics[width=\linewidth]{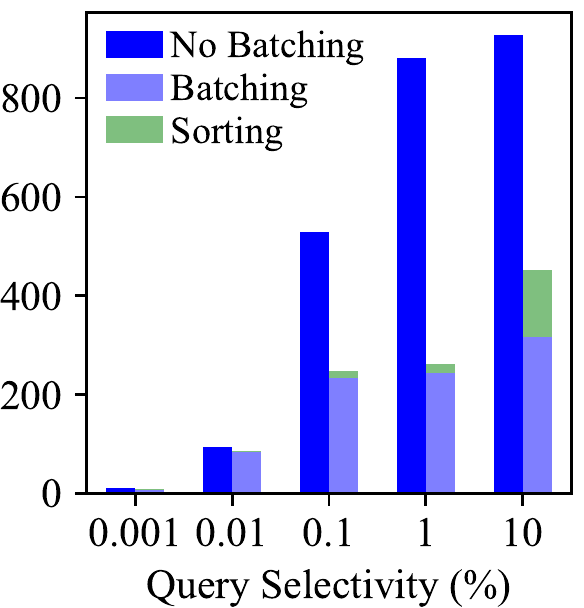}
		\caption{Impact of Sorting}
		\label{fig:query-batch-sort}
	\end{subfigure}%
	\caption{Effectiveness of Point Lookup Optimizations}
	\label{fig:query-opt}
\end{figure*}

\section{Experimental Evaluation}
\label{sec:evaluation}
In this section, we evaluate the proposed techniques in the context of Apache AsterixDB~\cite{asterixdb-web}.
We are primarily interested in evaluating the effectiveness of the various point lookup optimizations (\refsection{sec:expr-lookup})
and the ingestion and query performance of the Eager, Validation, and Mutable-bitmap strategies discussed in this paper (Sections \ref{sec:expr-ingestion-perf} - \ref{sec:expr-query-perf}).
We also evaluate the proposed secondary index repair method in detail (\refsection{sec:expr-index-repair}).
Finally, we examine the performance of the two proposed concurrency control methods for the Mutable-bitmap strategy (\refsection{sec:expr-concurrency-control}).

\subsection{Experimental Setup}
Since our work focuses on a partitioned database architecture,
all experiments were performed on a single node with a single dataset partition.
The overall performance of multiple partitions generally achieves near-linear speedup
since both data ingestion and query processing are performed at each partition locally.
The node has a 4-core AMD Opteron 2212 2.0GHZ CPU, 8GB of memory, and two 7200 rpm SATA hard disks.
We used one disk for transactional logging and one for the LSM storage.
We allocated 5GB of memory for the AsterixDB instance.
Within that allocation, the disk buffer cache size is set at 2GB, and each dataset is given a 128MB budget for its memory components.
Each LSM-tree has two memory components to minimize stalls during a flush.
We used a tiering merge policy with a size ratio of 1.2 throughout the experiments, similar to the one used in other systems~\cite{cassandra, hbase}.
This policy merges a sequence of components when the total size of the younger components is 1.2 times larger than that of the oldest component in the sequence.
We set the maximum mergeable component size at 1GB to simulate the effect of disk components accumulating within each experiment period.
Each LSM-tree is merged independently.
The Bloom filter false positive rate setting is 1\% and the data page size is set at 128KB to accommodate sequential I/Os.
Finally, whenever a scan is performed, a 4MB read-ahead size is used to minimize random I/Os.
To verify the applicability of the proposed techniques on SSDs, we repeated some key experiments on a node with 4-core Intel i7 2.70GHZ CPU, 16GB memory, and a 500GB SSD.
All configurations on the SSD node were the same, except that the buffer cache size is set to 4GB and disk page size is reduced to 32KB.

For the experimental workload, although YCSB~\cite{ycsb2010} is a popular benchmark for key-value store systems, it is not suitable for our evaluation
since it does not have secondary keys nor secondary index queries.
As a result, we implemented a synthetic tweet generator for our evaluation, as in other LSM secondary indexing work~\cite{asterixdb-storage2014,secondary2018}.
Each tweet has several attributes such as ID, message\_text, user\_id, location, and creation\_time etc.,
and its size is about 500 bytes with variations due to the variable length (450 to 550 bytes) of the randomly generated tweet messages.
Among these attributes, the following three are related to our evaluation.
First, each tweet has an ID as its primary key, which is a randomly generated 64-bit integer.
Second, the user\_id attribute, which is a randomly generated integer in the range 0 to 100K, is used for 
formulating secondary index queries with various controlled selectivities.
Finally, each tweet has a creation\_time attribute, which is a monotonically increasing timestamp used to test the range filter.

\subsection{Point Lookup Optimizations}
\label{sec:expr-lookup}
We first studied the effectiveness of the various point lookup optimizations discussed in \refsection{sec:index-navigation}.
To this end, we performed a detailed experimental analysis as follows:
On top of a naive lookup implementation (denoted as ``naive'') which only sorts the primary keys,
we enabled the batched point lookup with a batch size of 16MB (denoted as ``batch''),
stateful \btree lookup (denoted as ``sLookup''), and blocked Bloom filter (denoted as ``bBF'') optimizations one by one.
We also studied another optimization proposed by Jia~\cite{jia2017} (denoted as ``pID'').
Its basic idea is to propagate and use the component IDs of the secondary index components
in which keys are found to prune primary index components during point lookups\footnote{Jia's original technique actually propagated the ranges of the range filter built on a time-correlated attribute; this is equivalent to propagating component IDs in our setting.}.

To prepare the experiment dataset, we inserted 80 million tweet records with no updates.
The resulting primary index has about 30GB of data and 20 disk components, while the secondary index on the user\_id attribute only has 3GB of data and less than 10 disk components.
We then evaluated the query performance for different controlled selectivities based on the user\_id attribute.
For each query selectivity, queries with different range predicates were executed until the cache was warmed and the average stable running time is reported.
The running time of different selectivities is reported in Figures \ref{fig:query-lookup-opt-low} and \ref{fig:query-lookup-opt-high}.
For low query selectivities (\reffigure{fig:query-lookup-opt-low}), which select a small number of records, batching slightly improves query performance by reducing random I/Os,
while other optimizations are less effective since the query time is dominated by disk I/Os.
For high query selectivities (\reffigure{fig:query-lookup-opt-high}), we also included the full scan time as a baseline.
The upper line, at about 750s, represents the full scan time on the experimental dataset.
Since disk read-ahead caused long I/O wait times,
we further evaluated an optimized case on a dataset prepared with sequential primary keys (lower line).
In that case, sequential I/Os and the OS's asynchronous read-ahead were fully exploited to minimize the I/O wait time.
As the query selectivity increases, the running time of the naive point lookup implementation grows quickly,
since accessing multiple LSM components leads to random I/Os.
The batched point lookup is the most effective at avoiding random I/Os.
For large selectivities, the stateful \btree lookup and blocked Bloom filter optimizations
further reduce the in-memory search cost,
since disk I/O cost is bounded as most pages must be accessed.
Note that for large selectivities, a full scan starts to outperform a secondary index search
since most pages of the primary index must be accessed
while a secondary index search incurs the extra cost of accessing the secondary index and sorting primary keys.
However, these optimizations together greatly improved the range of applicability of LSM-based secondary indexes,
allowing query optimizers, especially rule-based ones, to have more confidence when choosing secondary indexes.
Contrary to Jia~\cite{jia2017}, we found here that propagating component IDs provides little benefit.
The reason is that Jia~\cite{jia2017} considered an append-only temporal workload with hundreds of (filtered) LSM components in a dataset,
so the propagation of component IDs led to the skipping of a large number of Bloom filter tests.
However, in a more general setting, skipping a smaller number of Bloom filter tests per key would not make so big a difference.

We further evaluated batched point lookup in terms of the available batching memory and sorting overhead.
In each experiment that follows, stateful \btree lookup and blocked Bloom filter were enabled by default.
The running time of different query selectivities under different batch sizes is shown in \reffigure{fig:query-batch-size}.
For selective queries, a small batch size such as 128KB already provides optimal performance
since a batch of keys are often distributed over a large number of pages;
for non-selective queries, a few megabytes suffice to provide optimal performance as well.
Finally, since batching destroys the primary key ordering of the final results, we further evaluated the sorting overhead
by either not using batching or using batching plus sorting.
The running time of different query selectivities with these two query plans is shown in \reffigure{fig:query-batch-sort}.
Even though sorting must be performed, batching still improves the overall query performance.
This is because the point lookup step needs to fetch records distributed across a large number of pages,
while the resulting records can often fit into a small number of pages and can be sorted efficiently.

To summarize, batched point lookup is the most effective optimization for reducing random I/Os when accessing LSM components.
Stateful \btree lookup and blocked Bloom filter
are mainly effective for non-selective queries at further reducing the in-memory search cost.

\begin{figure*}
		\begin{minipage}[t]{.42\linewidth}
		\centering
		\includegraphics[width=\linewidth]{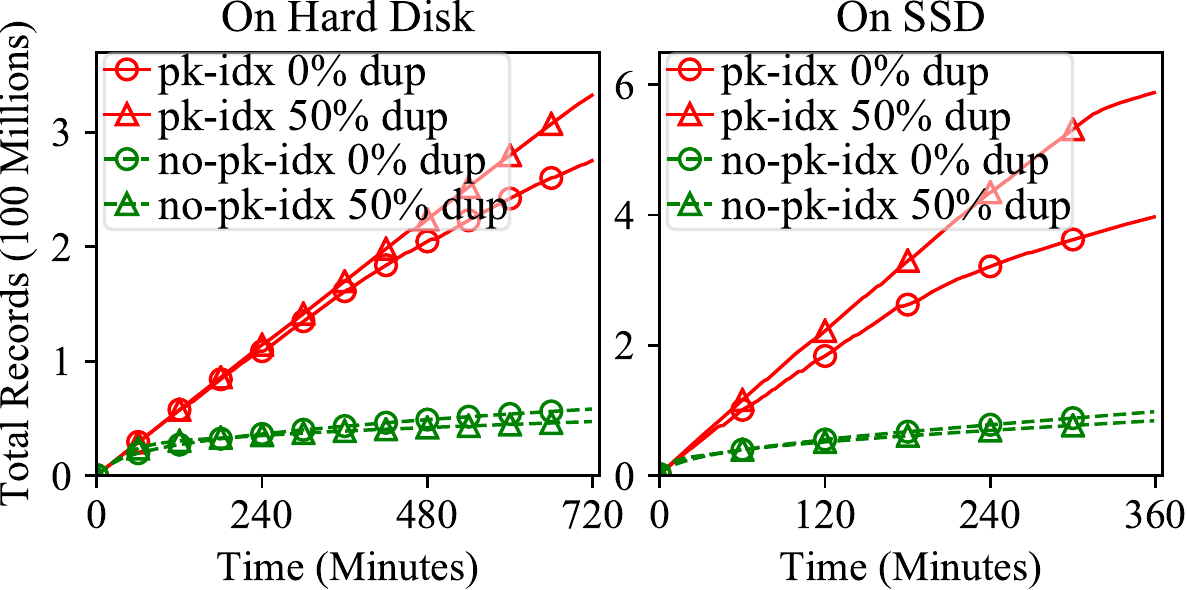}
		\caption{Insert Ingestion Performance}
		\label{fig:insert-expr}
	\end{minipage}%
	\begin{minipage}[t]{.58\linewidth}
		\centering
		\includegraphics[width=1\linewidth]{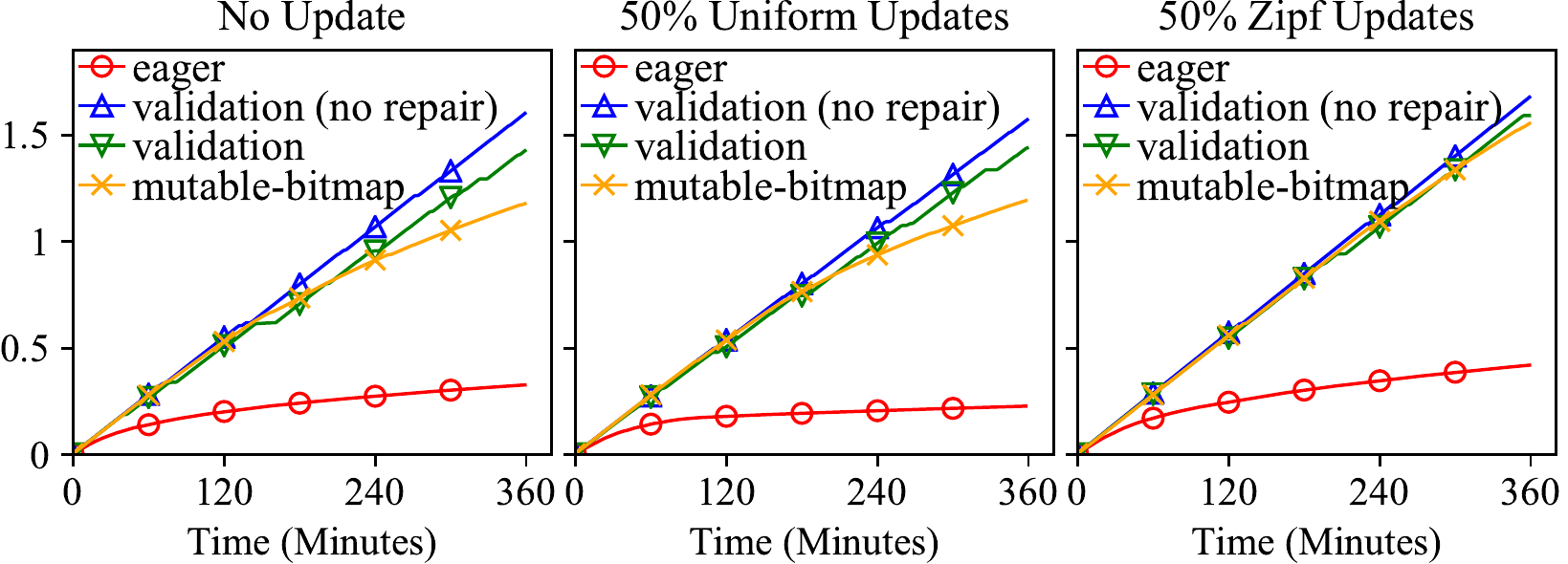}
		\caption{Upsert Ingestion Performance}
		\label{fig:upsert-expr-basic}
	\end{minipage}%
\end{figure*}

\subsection{Ingestion Performance}
\label{sec:expr-ingestion-perf}
We next evaluated the ingestion performance of the different maintenance strategies, focusing on the following key questions:
(1) What is the effectiveness of building a primary key index, which reduces the point lookup cost, for improving the overall ingestion throughput,
since maintaining the primary key index itself incurs extra cost?
(2) Does repairing secondary indexes become a new bottleneck, especially for a dataset with multiple secondary indexes?
(3) What is the effectiveness of building a single primary key index and the proposed repair algorithm (\refsection{sec:index-repair})
for improving the ingestion throughput as compared to the deleted-key \btree strategy supported by AsterixDB?

To answer these questions, we used insert and upsert workloads in the following evaluation.
Delete workloads were omitted since the cost of deletes would be similar to upserts; that is, an upsert is logically equivalent to a delete plus an insert.
Unless otherwise noted, each experiment below was run for 6 hours with the goal being to ingest as much data as possible without concurrent queries.
That is, each experiment used a 100\% write workload.
Each dataset has a primary index with a component-level range filter on the creation\_time attribute,
a primary key index, and a secondary index on the user\_id attribute.

\subsubsection{Insert Workload}
For inserts, we evaluated two methods for enforcing key uniqueness, using either the primary index or the primary key index.
In the former case, the primary key index was omitted to eliminate its overhead.
This workload was controlled by a duplicate ratio, which is the ratio of duplicates among all records.
Duplicates were randomly generated following a uniform distribution over all the past keys.

\reffigure{fig:insert-expr} shows the insert throughput under different duplicate ratios on both hard disks and SSDs.
Note that the experiment on hard disks was run for 12 hours so that even the primary key index cannot be totally cached.
As the result shows, without the primary key index, ingestion performance degraded quickly once the dataset could not be totally cached;
point lookups during ingestion incur a large overhead on both hard disks and SSDs.
Building a primary key index greatly improves the ingestion throughput since the keys are much smaller and can be better cached to reduce disk I/Os caused by point lookups.
Even though the ingestion throughput drops when the primary key index cannot be totally cached, building a primary key index is still helpful by increasing the cache hit ratio.
The duplicate-heavy workload results in higher ingestion throughput when using the primary key index since duplicate keys are simply excluded from insertion into the storage.
However, when the primary key index is not built, the duplicate-heavy workload reduces the throughput because of a large number of random I/Os for the uniqueness check.

\subsubsection{Upsert Workload}
\label{sec:upsert-ingestion}
The upsert workload is the main focus of our evaluation since the three maintenance strategies discussed in this paper mainly differ in how upserts are handled.
This workload was controlled by an update ratio, which is the ratio of updates (records with past ingested keys) among total records.
Updates are randomly generated by following either a uniform distribution, that is, all past keys are updated equally,
or a Zipf distribution with a theta value 0.99 as in the YCSB benchmark~\cite{ycsb2010}, that is, recently ingested keys are updated more frequently.
Unless otherwise specified, the update ratio was chosen as 10\% and the updates followed a uniform distribution.

For the Validation strategy, we evaluated two variations to measure the overhead of repairing secondary indexes.
In the first variation, repair was totally disabled to maximize ingestion performance.
In the second variation, merge repair was enabled but without the Bloom filter optimization (\refsection{sec:index-repair})
to evaluate the worse case repairing overhead.
For the Mutable-bitmap strategy, the secondary index was maintained using the Validation strategy without repair to minimize the ingestion overhead due to secondary indexes.
The Side-file method was used for concurrency control to minimize the locking overhead.

\textbf{Basic Upsert Ingestion Performance.}
In this experiment, we compared the proposed Validation and Mutable-bitmap strategies with the Eager strategy
in terms of the upsert ingestion performance under different update ratios,
ranging from no updates to 50\% updates following either the uniform or Zipf distribution.
The experimental results are shown in \reffigure{fig:upsert-expr-basic}.
The Eager strategy, which ensures that secondary indexes are always up-to-date, has the worst ingestion performance
because of the point lookups to maintain secondary indexes using the old records.
On the other hand, the Validation strategy without repairing has the best ingestion performance since secondary indexes are not cleaned up at all.
The result also shows that repairing secondary indexes incurs only a small amount of extra overhead.
Of course, since secondary indexes are not always up-to-date, the Validation strategy sacrifices query performance,
which we will evaluate later in \refsection{sec:expr-query-perf}. 
The Mutable-bitmap strategy also has much better ingestion performance than the Eager strategy
since it only searches the primary key index instead of accessing full records.
Updates generally have a small impact on the overall ingestion throughput, since updated records must be inserted into memory and subsequently flushed and merged (as for inserted records).
The Eager and Mutable-bitmap strategies both benefit from skewed update workloads since most of the updates only touch recent keys, reducing disk I/O incurred by the point lookups.
These results again confirm that it is helpful to build a primary key index to reduce the point lookup cost.

\begin{figure}[b]
	\centering
	\includegraphics[width=0.8\linewidth]{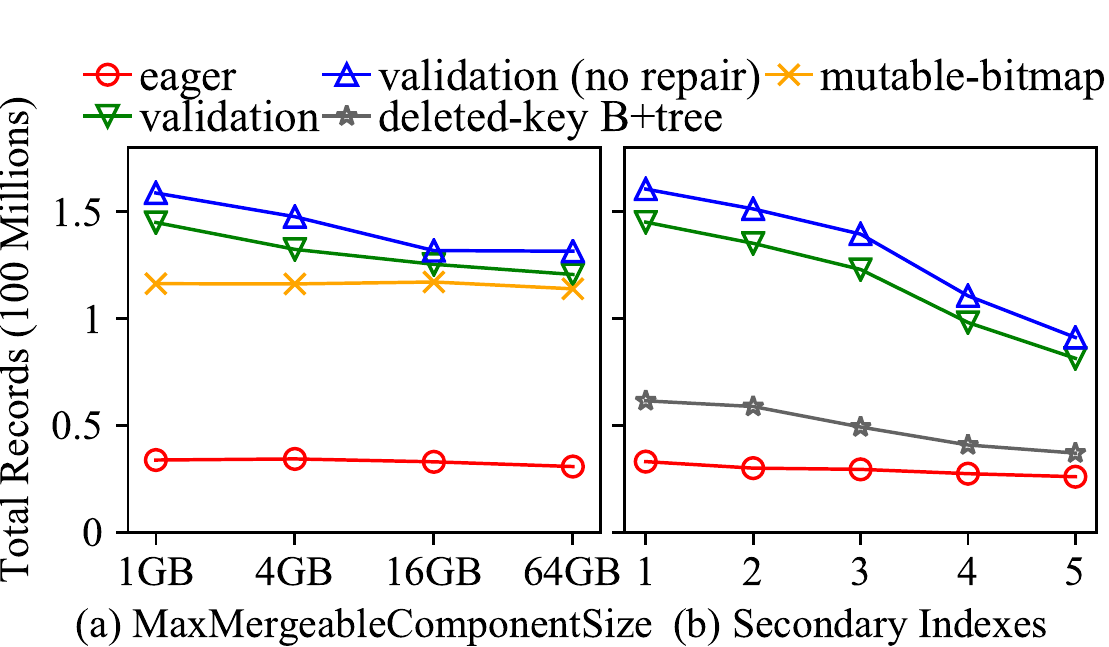}
	\caption{Impact of Merge Operations and Secondary Indexes on Upsert Ingestion Performance}
	\label{fig:upsert-merge-index}
\end{figure}%

\begin{figure*}
	\begin{minipage}[t]{.541\linewidth}
		\centering
		\includegraphics[width=\linewidth]{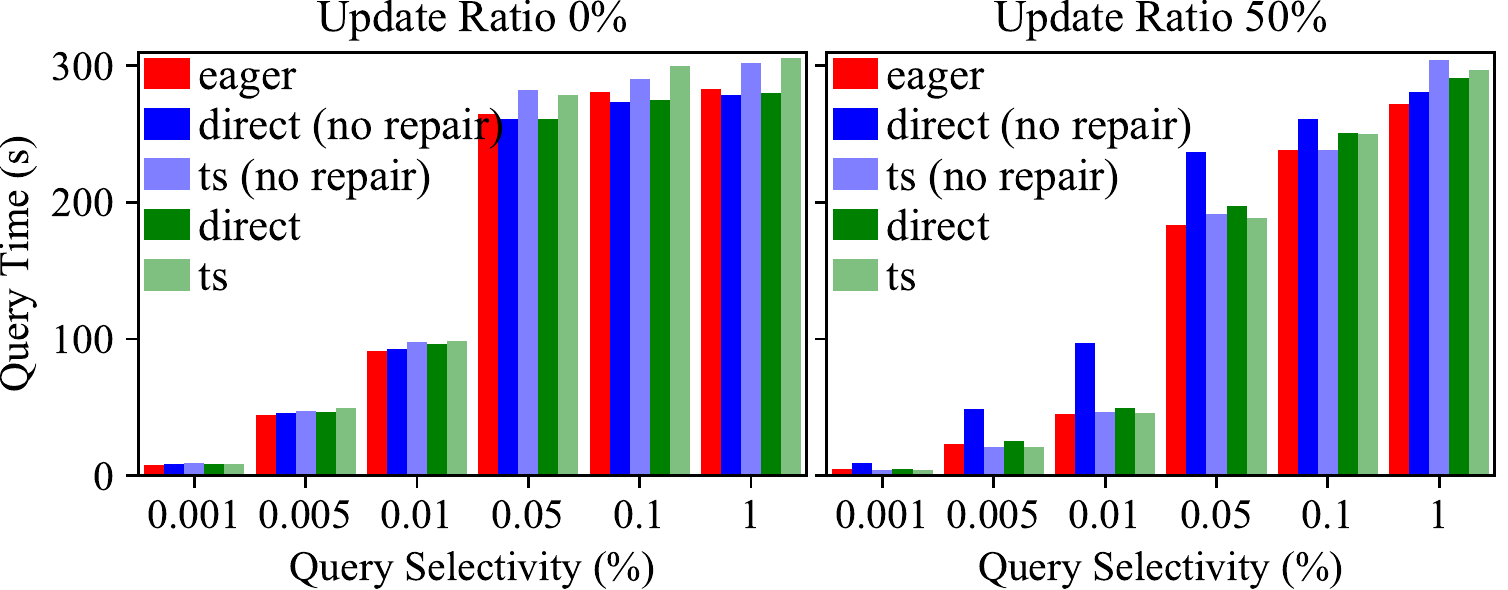}
		\caption{Non-Index-Only Query Performance}
		\label{fig:query-secondary-index}
	\end{minipage}%
	\begin{minipage}[t]{0.449\linewidth}
		\centering
		\includegraphics[width=\linewidth]{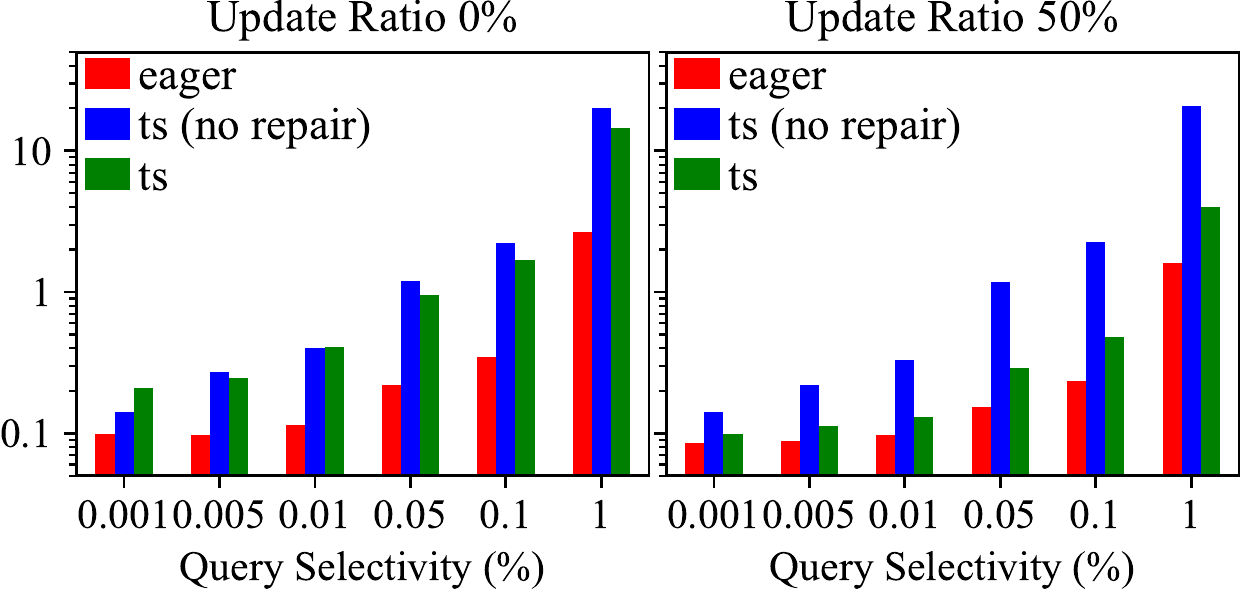}
		\caption{Index-Only Query Performance}
		\label{fig:query-secondary-index-only}
	\end{minipage}%
\end{figure*}

\textbf{Ingestion Impact of Merge Operations.}
In this set of experiments, we evaluated the impact of merge operations on upsert ingestion performance
since more merges can reduce the point lookup cost by reducing the number of components.
To evaluate this, we varied the maximum mergeable component size to control the merge frequency.
The resulting ingestion throughput is shown in \reffigure{fig:upsert-merge-index}a.
Except that more merges negatively impact the ingestion throughput of all strategies,
point lookups still incur a large amount of overhead during data ingestion and the same relative performance trends still hold.

\textbf{Ingestion Impact of Secondary Indexes.}
We also evaluated the scalability of the strategies by adding more secondary indexes.
We excluded the Mutable-bitmap strategy, since it is unaffected by secondary indexes,
but instead included the deleted-key \btree strategy for comparison.
\reffigure{fig:upsert-merge-index}b shows the ingestion throughput for different numbers of secondary indexes.
Maintaining more secondary indexes negatively impacts the ingestion performance of all the strategies since more LSM-trees must be maintained.
It also has a larger impact on the Validation strategies.
The reason is that the bottleneck of the Eager strategy lies in the point lookups, while for the Validation strategy, its bottleneck is flushing and merging the LSM-trees.
This impact is more obvious after adding the fourth secondary index, since our experiment machine has four cores and
more secondary indexes result in higher contention among all merge threads.
The experiment also shows the scalability of the proposed index repair operations since it introduces just a small overhead on data ingestion.
Comparing to the current AsterixDB deleted-key \btree strategy,
the negative impact of index repair operations has been greatly reduced by using a single primary key index 
and the efficient repair algorithm presented in \refsection{sec:index-repair}.

\subsection{Query Performance}
\label{sec:expr-query-perf}
We next evaluated the query performance of the different maintenance strategies, focusing on the following two aspects.
First, we evaluated the overhead of the Validation strategy with both non-index-only and index-only
queries compared to the read-optimized Eager strategy and the benefit of repairing secondary indexes.
Second, we evaluated the pruning capabilities of filters resulting from different maintenance strategies.
As in \refsection{sec:expr-lookup}, each dataset that follows was prepared by upserting 80 million records
with different actual update ratios (0\% or 50\%).

\subsubsection{Secondary Index Query Performance}
We first evaluated the overhead of the Validation strategy on secondary index range queries.
We again considered two variations of the Validation strategy, depending on whether merge repair was enabled, to evaluate the benefit of repairing secondary indexes.
Each query for a given selectivity was repeated with different range predicates until the cache was warmed and the average stable time is reported.

\begin{figure}[t]
	\centering
	\includegraphics[width=0.65\linewidth]{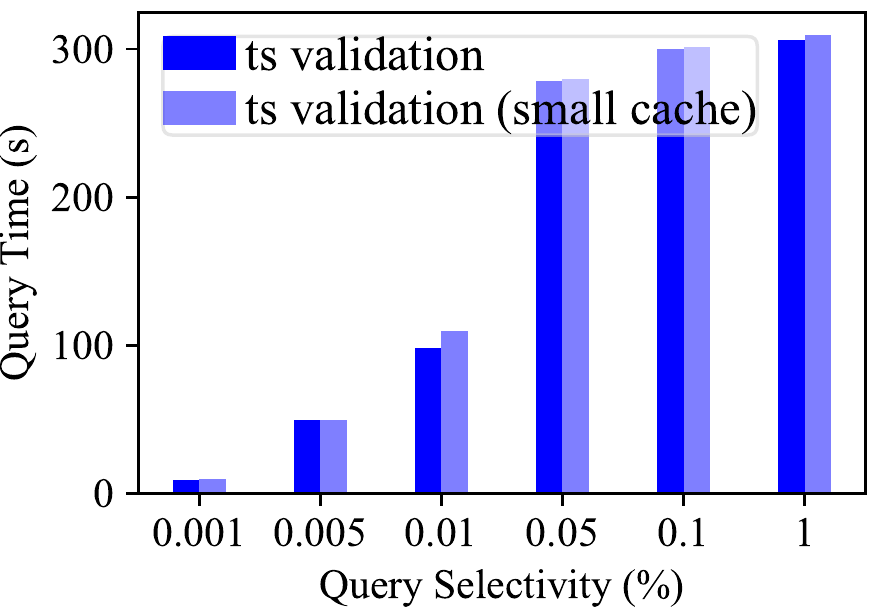}
	\caption{Impact of Small Cache on Timestamp Validation}
	\label{fig:query-secondary-index-small-cache}
\end{figure}

\textbf{Non-Index-Only Query Performance.}
For non-index-only queries, we enabled batching (with 16MB memory), stateful \btree search, and blocked Bloom filter
to optimize the subsequent point lookups.
For the Validation strategy, we evaluated both Direct Validation (denoted as ``direct'') and Timestamp Validation (denoted as ``ts'').
The running time of non-index-only queries is shown in \reffigure{fig:query-secondary-index}.
In general, the Validation strategy has comparable performance to the Eager strategy for non-index-only queries.
For the append-only workload (left side), the Direct Validation method has similar performance to the Eager strategy,
since secondary indexes do not contain any obsolete entries.
However, the Timestamp Validation method leads to some extra overhead because of the validation step.
The results are different for the update-heavy workload (right side).
When the secondary index is not repaired, the Direct Validation method leads to high overhead
for selective queries, as a lot of I/O is wasted searching for obsolete keys.
However, the extra overhead diminishes when the selectivity becomes larger, such as 0.1 - 1\%,
since even searches for valid keys must access almost all pages of the primary index.
The Timestamp Validation method is helpful for reducing wasted I/O by filtering out the obsolete keys using the primary key index.
With merge repair, most of the obsolete keys would be removed,
and thus both validation methods would have comparable query performance to the Eager strategy.

We further evaluated the performance of the Timestamp Validation method under a smaller cache memory setting (only 512MB) so that the primary key index cannot totally fit into memory.
The dataset used in this experiment contains no updates.
\reffigure{fig:query-secondary-index-small-cache} depicts the query time of this experiment.
In general, we found that a small cache setting has limited impact on the Timestamp Validation method
since the primary key index is much smaller than the primary index,
resulting in only a small amount of extra I/Os during validation.

\textbf{Index-Only Query Performance.}
\reffigure{fig:query-secondary-index-only} shows the query time (log scale) for index-only queries.
We omit the Direct Validation method since it must fetch records for validation and thus has same performance as for non-index-only queries.
In general, the Validation strategy performs worse (3x - 5x) than the Eager strategy
because of its extra sorting and validation steps.
Even without obsolete entries (left side), the validation step still leads to extra overhead.
Still, Timestamp Validation provides much better performance than Direct Validation would
by accessing the primary key index.
Finally, merge repair is helpful for index-only queries by increasing repaired timestamps to prune more primary key index components during validation,
as shown in the append-only case (left side), and by cleaning up obsolete entries for the update-heavy case (right side).

\begin{figure}[b]
	\centering
	\includegraphics[width=\linewidth]{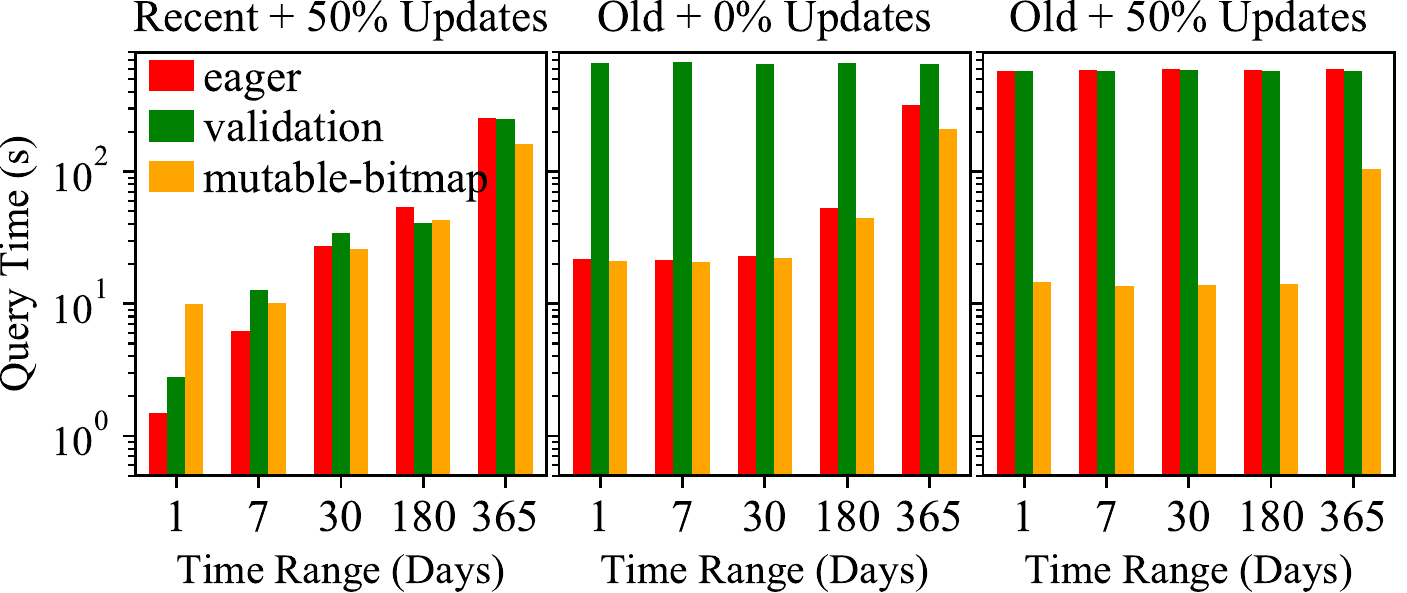}
	\caption{Query Performance of Range Filters}
	\label{fig:query-filter}
\end{figure}

\subsubsection{Range Filter Query Performance}
We next evaluated the effectiveness of the maintenance strategies for filters.
In particular, we used the range filter in our evaluation.
Recall that the range filter was built on the creation\_time attribute, which is a monotonically increasing timestamp.
The range of the creation\_time attribute of all tweet records in the experiment dataset spanned about 2 years.
Each query in this experiment has a range predicate on the creation\_time attribute,
and it is processed by scanning the primary index with the component-level pruning provided by the range filters.
We evaluated two types of queries, queries that access recent data  (with creation\_time $>$ T) and that access old data (with creation\_time $<$ T).
Each query was repeated 5 times with a clean cache for each run, and the average query time is reported.

The query times with range filters are summarized in \reffigure{fig:query-filter}.
For the queries that access recent data, all the strategies provide effective pruning capabilities.
The Mutable-bitmap strategy further improves scan performance since LSM components are accessed one by one and the reconciliation step is no longer needed.
In contrast, when accessing old data, the Validation strategy provides no pruning capability
since all newer components must be accessed to ensure correctness.
The Eager strategy is only effective for the append-only case, but recall that its point lookups lead to high cost during data ingestion.
The Mutable-bitmap strategy provides effective pruning capabilities via the use of Mutable bitmaps under all settings,
and does so with only a small amount of overhead on data ingestion.

\subsection{Index Repair Performance}
\label{sec:expr-index-repair}
We then evaluated the index repair performance of the proposed Validation strategy (referred as \emph{secondary repair}) in detail,
as well as the proposed Bloom filter optimization (denoted as ``bf'').
For comparison, we also evaluated the index repair method proposed by DELI~\cite{deli2015} (referred as \emph{primary repair}).
Recall that DELI repairs secondary indexes by merging or scanning primary index components to identify obsolete records,
while our approach uses a primary key index to avoid accessing full records.

In each of the following experiments, we upserted 100 million tweet records with merge repair enabled.
Since the two methods trigger repair operations differently, the resulting secondary indexes may have different amounts of obsolete entries during data ingestion.
To enable a fair comparison, instead of measuring the overall ingestion throughput, we measured the index repair performance directly as follows.
For every 10 million records ingested, we stopped the ingestion and triggered a full repair operation to bring all secondary indexes up-to-date.
This provides the trend of index repair performance as data accumulates.
For our secondary repair method, standalone repair was performed to exclude the extra overhead due to merges.

\begin{figure}[b]
	\centering
	\includegraphics[width=\linewidth]{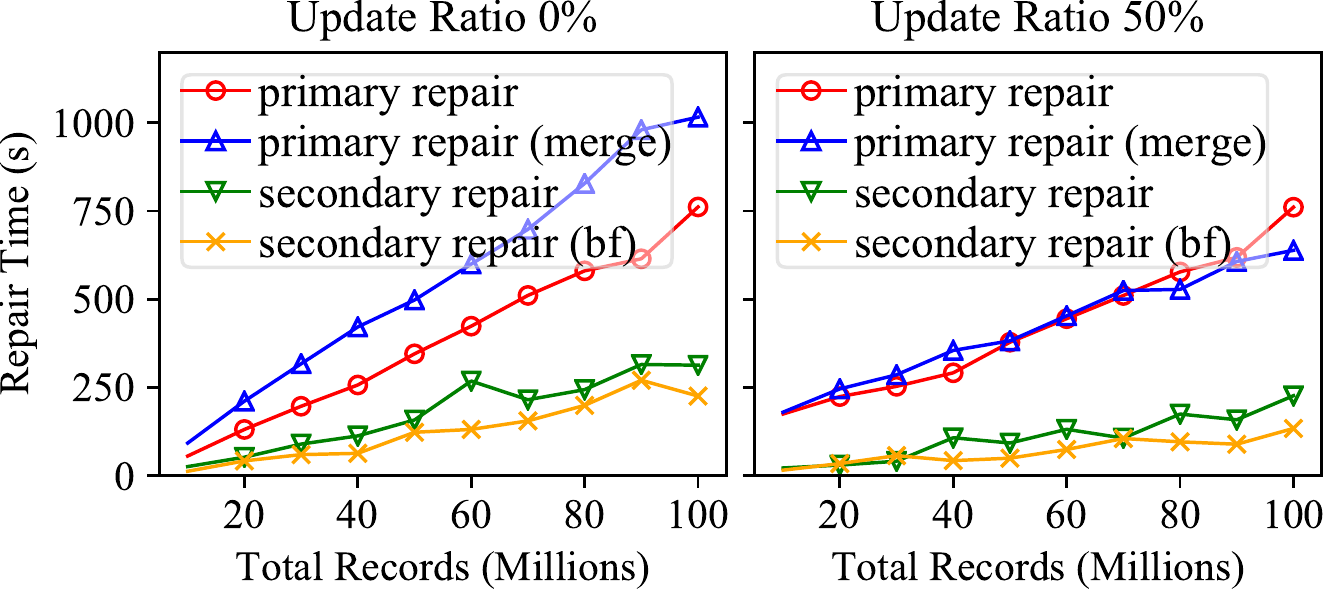}
	\caption{Repair Performance with Varying Update Ratio}
	\label{fig:repair-update}
\end{figure}

	\begin{figure}[t]
		\centering
		\includegraphics[width=0.65\linewidth]{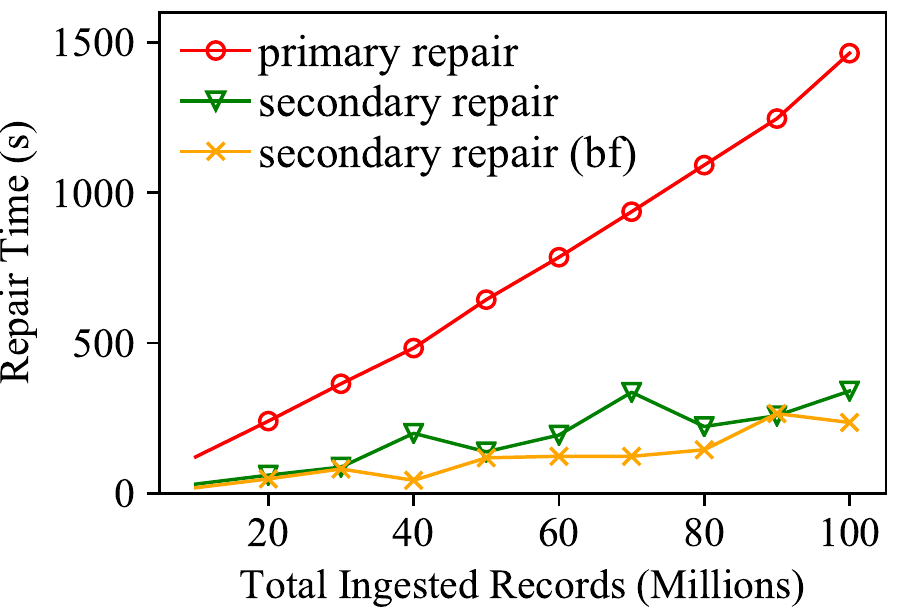}
		\caption{Repair Impact of Large Records}
		\label{fig:repair-record-size}
	\end{figure}

\begin{figure}
	\centering
	\includegraphics[width=0.6\linewidth]{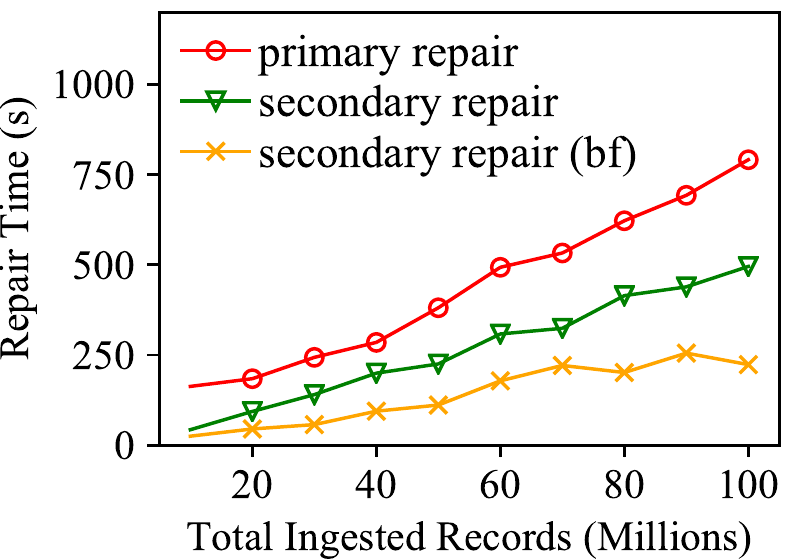}
	\caption{Repair Impact of Secondary Indexes} 
	\label{fig:repair-index}
\end{figure}

\textbf{Basic Repair Performance.}
We first evaluated the index repair performance under different update ratios (0\% and 50\%).
For primary repair~\cite{deli2015}, we evaluated two variations depending whether a full merge is performed as a by-product of the repair operation.
The index repair performance over time is depicted in \reffigure{fig:repair-update},
where each data point represents the time to complete a repair operation.
For primary repair, a full merge leads to extra overhead for append-only workloads,
but improves subsequent repair performance for update-heavy workloads.
The secondary repair method proposed here always outperforms the primary repair methods
because only it only accesses the primary key index, significantly reducing disk I/O.
The Bloom filter optimization further improves repair performance
by reducing the volume of primary keys to be sorted and validated.

\textbf{Impact of Large Records.}
In this set of experiments, we used large record size 1KB to evaluate the impact of large records.
The result is depicted in \reffigure{fig:repair-record-size}, where the update ratio is set to 10\%.
Large records negatively impact the primary repair method, since more I/O must be performed.
However, it has no impact over the secondary repair method, since it only accesses the primary key index.

\begin{figure*}[t]
	\centering
	\begin{subfigure}{0.28\textwidth}
		\centering
		\includegraphics[width=\linewidth]{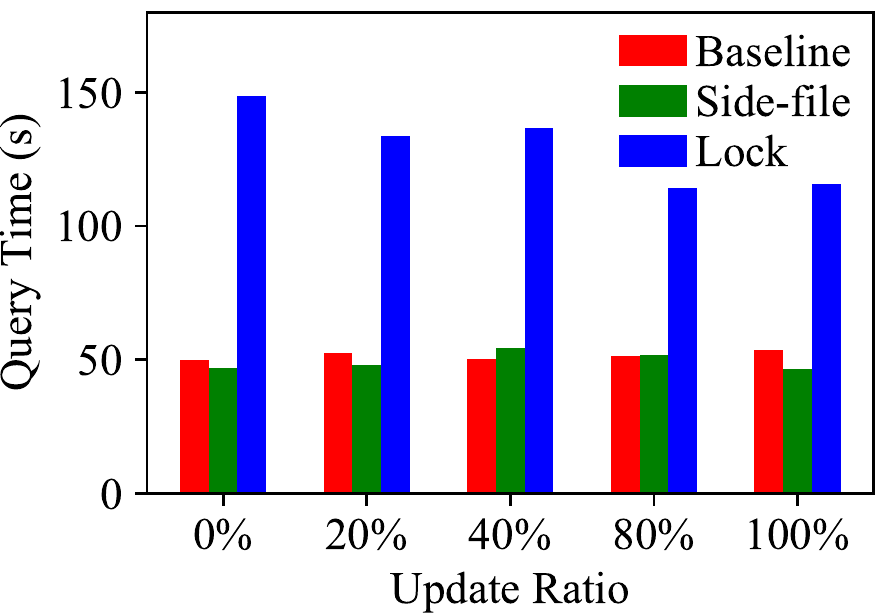}
		\caption{Impact of Updates}
		\label{fig:expr-bitmap-updates}
	\end{subfigure}%
	\hfil
	\begin{subfigure}{0.28\textwidth}
		\centering
		\includegraphics[width=\linewidth]{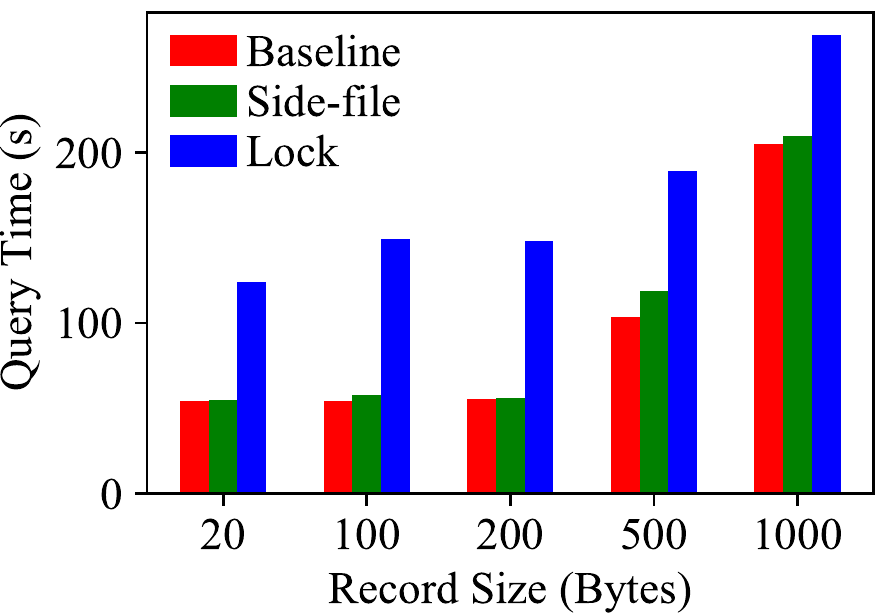}
		\caption{Impact of Component Size}
		\label{fig:expr-bitmap-component-sizes}
	\end{subfigure}
    \hfil
	\begin{subfigure}{0.28\textwidth}
		\centering
		\includegraphics[width=\linewidth]{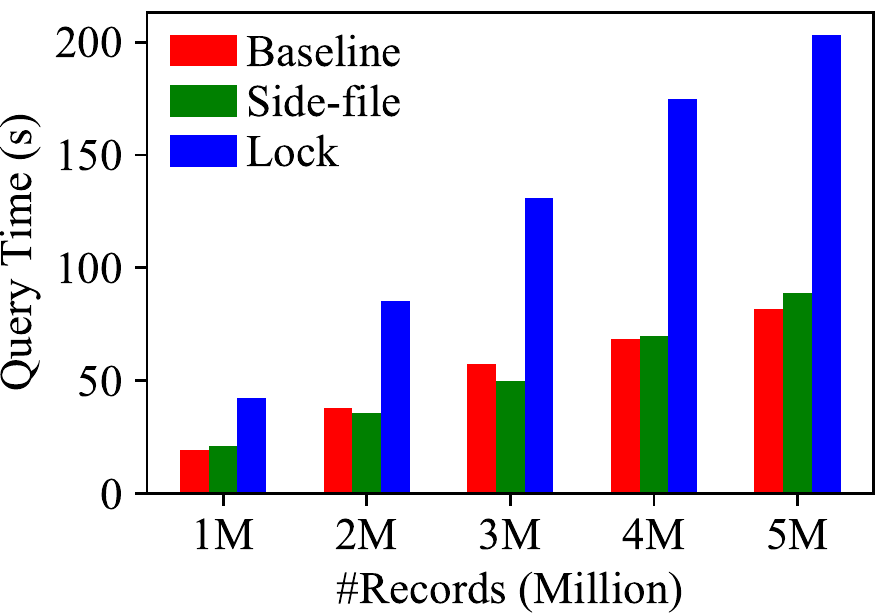}
		\caption{Impact of Record Size}
		\label{fig:expr-bitmap-record-sizes}
	\end{subfigure}%
	\caption{Overhead of Mutable-Bitmap Concurrency Control Methods}
	\label{fig:concurrency-control}
\end{figure*}

\textbf{Impact of Secondary Indexes.}
We next used 5 secondary indexes to evaluate the scalability of the index repair methods.
The update ratio was set to 10\%.
For the secondary repair method, each secondary index was repaired in parallel using multi-threading.
The index repair time is depicted in \reffigure{fig:repair-index}.
Having more secondary indexes negatively impacts the performance of both methods.
For the primary repair method, more anti-matter entries must be inserted into more secondary indexes.
For the secondary repair method, more secondary index entries must be scanned and sorted for validation.
The result does show that the proposed secondary index repair method is easily parallelizable,
since it only performs a small amount of I/O and most operations are CPU-heavy.
Furthermore, the Bloom filter optimization reduces the negative impact of having more secondary indexes,
as it sorts fewer keys, significantly reducing the I/O overhead during the sort step.

\subsection{Concurrency Control for Mutable-Bitmap}
\label{sec:expr-concurrency-control}
In the final set of experiments, we evaluated the overhead of the alternative concurrency control methods proposed
for the Mutable-bitmap strategy, that is, the Lock method and the Side-file method.
We chose the merge time without any concurrency control as the baseline.
In each experiment, we merged 4 components with concurrent data ingestion at the maximum speed.
Unless otherwise noted, each component had 3 million records, each record had 100 bytes,
and the update ratio of the newly ingested records was set to 50\%.

We evaluated the impact of update ratio, the size of merged components, and the record size 
on the performance of the proposed concurrency control methods.
The experiment results are summarized in \reffigure{fig:concurrency-control},
In general, the Side-file method incurs negligible overhead against the baseline
since no locks are acquired on a record-basis.
In contrast, the Lock method performs worse in all the settings because of the locking overhead,
even though the locking overhead is marginalized as the record size grows larger.
The Lock method also benefits from updates,
since the updated entries are simply skipped during the merge, while the Side-file method has to apply updates later.

\section{Conclusion}
\label{sec:conclusion}
In this paper, we have presented techniques for efficient data ingestion and query processing for LSM-based storage systems.
We described and evaluated a series of optimizations for efficient point lookups, greatly improving the range of applicability for LSM-based secondary indexes.
We further presented new and efficient strategies for maintaining LSM-based auxiliary structures, including secondary indexes and filters.
The Validation strategy defers secondary index maintenance to the background using a primary key index,
significantly improving ingestion performance by eliminating ingestion-time point lookups.
This leads to a small overhead for non-index-only queries, but a relatively high overhead for index-only queries because of the extra validation step.
The Mutable-bitmap strategy maximizes the pruning capabilities of filters
and improves ingestion performance by only accessing the primary key index, instead of full records, during data ingestion.

We plan to extend our work in several directions.
First, we hope to extend the Validation strategy to support more efficient processing of index-only queries, further improving its applicability.
Second, we plan extend the proposed strategies to let queries drive the maintenance of auxiliary structures, as suggested by database cracking~\cite{cracking2007}.
Finally, since no strategy was found to work best for all workloads,
we plan to develop auto-tuning techniques so that the system could dynamically adopt the optimal maintenance strategies for a given workload.

\noindent
\textbf{Acknowledgments.}
This work is supported by NSF awards CNS-1305430, IIS-1447720, and IIS-1838248 along with industrial support from Amazon, Google, and Microsoft and support from the Donald Bren Foundation (via a Bren Chair).

\balance

\bibliographystyle{abbrv}
\bibliography{index}
	
\end{document}